\documentclass[a4paper, 11pt]{article}
\usepackage[pdftex, a4paper]{geometry}
\usepackage{graphicx}				
				
\usepackage{amsmath}
\usepackage{amsfonts}
\usepackage{dsfont}
\usepackage{mathrsfs}
\usepackage{amssymb}
\usepackage{bbm}
\usepackage{epsfig}
\usepackage[english]{babel}
\usepackage[all]{xy}

\def\T{^{\rm\tiny T}}

\newtheorem{theorem}{Theorem}
\newtheorem{definition}{Definition}

\newtheorem{lemma}{Lemma}
\newtheorem{remark}{Remark}

\begin{document}

\title{\bf Reaching Quantum  Consensus with Directed Links:     Missing Symmetry and Switching Interactions\footnote{A preliminary version of this work will be presented at the American Control Conference, Chicago, July 1--3, 2015 \cite{ACC1,ACC2}.  G. Shi and S. Fu  are with the Research School of Engineering, The Australian National University, Canberra 0200, Australia. I. R. Petersen is with School of Engineering and Information Technology, University of New South Wales, Canberra 2600, Australia. Email: guodong.shi, shuangshuang.fu@anu.edu.au,  i.r.petersen@gmail.com.} }
\date{}

\author{Guodong Shi, Shuangshuang Fu, and Ian R. Petersen}
\maketitle

\begin{abstract}
In this paper, we study  consensus  seeking of quantum  networks  under directed interactions defined  by a set of permutation operators among a network of qubits. The state evolution of the quantum network is described by a continuous-time master equation, for which we establish an unconditional convergence result indicating that the network state always converges with the limit determined by the generating subgroup of the permutations making use of the Perron-Frobenius theory. We also give a tight graphical criterion regarding when such limit admits a reduced-state consensus.  Further, we provide a  clear description  to the missing symmetry in the reduced-state consensus from a graphical point of view, where  the information-flow  hierarchy in quantum permutation operators is characterized by  different layers of information-induced graphs. Finally, we investigate   quantum synchronization in the presence of network Hamiltonian, study  quantum consensus conditions under switching interactions, and present a few numerical examples illustrating the obtained results.
\end{abstract}


\section{Introduction}
\subsection{Background and related works}
Consensus and synchronization of node states in a network of coupled dynamics have been extensively studied in the past decades  \cite{WU-CHUA95,jad03,Jadbabaie2004}.
The basic idea is that  \cite{tsi} nodes in a network with suitable connectivity  can reach a common state or trajectory where all nodes' initial values are encoded with distributed node interactions, and the strong involvement of graph-theoretic  approaches   has  reshaped the study of networked control systems  \cite{Magnus}. Excellent results have been derived towards the understandings of   how bidirectional or directed, fixed or switching, deterministic or random, node interactions influence the consensus value and the speed  of consensus, e.g., \cite{saber04,Ren05,Hatao2005,Olshevsky2009}.

Scientific interest on consensus and synchronization subject to the laws of  quantum mechanics  has also been noticed.  Recent work \cite{PRL2013} introduced the measures of synchronization in quantum systems of coupled Harmonic oscillators. Sepulchre \emph{et al.}  \cite{Sepulchre-non-communtative} generalized consensus algorithms to non-commutative spaces and presented convergence results for quantum stochastic maps  to a fully mixed state.  Mazzarella \emph{et al.} \cite{Ticozzi} made a systematic study of  consensus-seeking in quantum networks, where  four classes of consensus quantum states based on invariance and symmetry properties were introduced and  a quantum gossip algorithm \cite{Boyd-gossip} was proposed for reaching a symmetric consensus state over a  quantum network. The class of quantum gossip algorithms was extended to  symmetrization problems in a group-theoretic framework  in \cite{Ticozzi-SIAM}.


Developments in continuous-time quantum consensus seeking were made  in \cite{Ticozzi-MTNS,ShiWCICA,ShiDongPetersenJohansson} for Markovian quantum dynamics governed by  master equations \cite{Breuer and Petruccione 2002}. In \cite{Ticozzi-MTNS}, using a group-theoretic analysis,  the authors proposed a consensus master equation involving  permutation operators and showed that a symmetric state consensus can be achieved under such evolutions. In \cite{ShiDongPetersenJohansson}, a graphical method was systematically introduced for building the connection between the quantum consensus dynamics and its classical analogue by  splitting  the evolution of the entries in the network density operator.  The idea of breaking down large density operators of multiple qubits can in fact be traced back to \cite{Altafini} using Stokes tensors.  A type of  quantum synchronization was also shown in the sense that the network trajectory tends to an orbit determined by the network Hamiltonian and the symmetrization of the initial state, which implies that   all quantum nodes asymptotically reach the same orbit  \cite{ShiDongPetersenJohansson}. For research on quantum network control and information processing we refer to \cite{NaturePhysiscs,Wang et al 2012,Gough and James 2009}; for a survey for quantum control theory we refer to \cite{Altafini and Ticozzi 2012}.
\subsection{Our contributions}
In this paper, we aim to present a thorough   investigation of consensus of  qubit (i.e., quantum bit) networks with directed qubit interactions. The evolution of the quantum network state is given by a Lindblad master equation  where the Lindblad operators are in a set of qubit permutations \cite{Ticozzi-MTNS}. We define a directed quantum interaction graph associated with  each of the permutation operators. Our first contribution is the establishment of an unconditional convergence result indicating that the network state always converges, and  the convergence limit is determined by the generating subgroup of the permutations. This result is proved via the Perron-Frobenius theory for non-negative matrices,  and our analysis also allows for characterization of the convergence speed. Moreover, we establish  a tight graphical criterion regarding when such limit admits a consensus in the qubits' reduced states.

Next, we  study the missing symmetry in the reduced-state  consensus state, compared to the symmetric consensus state studied  in \cite{Ticozzi,Ticozzi-MTNS}. The tool is  based on  extension of the graph-theoretic  methods used in \cite{ShiDongPetersenJohansson} to the directed case, and    the information-flow  hierarchy in quantum permutation operators is precisely captured by  different layers of information-induced graphs. Such a characterization  has proven able to  establish a clear bridge between quantum and classical consensus dynamics, based on which the full details in the quantum state evolution can be  visualized (cf., \cite{ShiDongPetersenJohansson}).

Finally, we  study synchronization of quantum networks with presence of network Hamiltonian as well as quantum consensus seeking subject to switching interactions in lights of the previously  obtained results. Particularly, we show that the quantum consensus dynamics with switching interactions  are by nature equivalent to a parallel  cut-balanced classical consensus processes \cite{juliencut}, and then a necessary and sufficient condition is obtained for the convergence of the dynamics under switching interactions.
\subsection{Paper organization}
The remainder of this paper is organized as follows. Section~\ref{Sec2} presents some preliminaries including
relevant concepts in linear algebra, graph theory and quantum systems. Section~\ref{Secproblem} introduces  the n-qubits network
model, its state evolution and the problem of interest. Section~\ref{SecConvergence} establishes the convergence condition  for the considered quantum network. Section~\ref{SecMissing} turns to a graph-theoretic  description of the information hierarchy in the qubit permutations, based on which we obtain a full interpretation of the missing symmetry between reduced-state consensus and symmetric-state consensus. Section~\ref{SecDiscussion} further discusses switching qubit interactions and quantum synchronization as well as presents a few numerical examples.  Finally, Section~\ref{SecConclusion} concludes the paper with a few remarks.

\section{Preliminaries}\label{Sec2}
In this section, we introduce some concepts and theories in linear algebra \cite{Horn}, graph theory \cite{godsil}, and quantum systems \cite{Nielsen}.

\subsection{Directed graphs}
A  (simple) directed graph  $\mathrm
{G}=(\mathrm {V}, \mathrm {E})$, or in short, a digraph,  consists of a finite set
$\mathrm{V}=\{1,\dots,N\}$ of nodes and an arc set
$\mathrm {E}$, where  an element $e=(i,j)\in\mathrm {E}$ denotes   an
{\it arc} from node $i\in \mathrm{V}$  to $j\in\mathrm{V}$ with $i\neq j$.  A directed path between two vertices $v_1$ and $v_k$ in $\mathrm{G}$ is a sequence of distinct nodes
$
v_1v_2\dots v_{k}$
such that for any $m=1,\dots,k-1$, there is an arc from $v_m$ to $v_{m+1}$; $
v_1v_2\dots v_{k}$ is called a semi-path if for any $m=1,\dots,k-1$, either $(v_m, v_{m+1})\in\mathrm{E}$ or $(v_{m+1}, v_{m})\in\mathrm{E}$. We call graph $\mathrm{G}$ to be {\it fully connected} if $(i,j)\in \mathrm{E}$ for all $i\neq j\in\mathsf{V}$; {\it strongly  connected} if, for every pair of distinct nodes in $\mathrm{V}$, there is a path  from one to the other;  {\it quasi-strongly connected} if there exists a node $v\in\mathcal{V}$ such that there is a path from $v$ to all other nodes; {\it weakly connected} if there is a semi-path between any two distinct nodes. The in-degree of $v\in \mathrm{V}$, denoted $\deg^-(v)$, is the number of nodes from which there is an arc entering $v$. The out-degree $\deg^+(v)$ can be correspondingly defined. The directed graph $\mathrm
{G}$ is called {\it balanced} if $\deg^+(v)=\deg^-(v)$ for any $v\in\mathrm{V}$.

A  subgraph of $\mathrm{G}$ associated with  $\mathrm{V}^\ast \subseteq \mathrm{V}$, denoted $\mathrm{G}|_{\mathrm{V}^\ast}$, is the graph $(\mathrm{V}^\ast, \mathrm{E}^\ast)$ with $(i,j)\in \mathrm{E}^\ast$ if and only if $(i,j)\in \mathrm{E}$ for $i,j\in \mathrm{V}^\ast$.  A {\it weakly connected component} (or, simply {\it component}) of  $\mathrm
{G}$ is a weakly connected subgraph associated with some $\mathrm{V}^\ast \subseteq \mathrm{V}$, with no  arc   between $\mathrm{V}^\ast$ and $\mathrm
{V}\setminus \mathrm{V}^\ast$.  The following lemma  characterizes   the connectivity of balanced digraphs. We provide a simple proof in  Appendix A.1.

\medskip

\begin{lemma}\label{lemmabalance}
A balanced digraph $\mathrm
{G}=(\mathrm {V}, \mathrm {E})$ is weakly connected if and only if it is strongly connected.
\end{lemma}

The Laplacian of $\mathrm
{G}$, denoted $L(\mathrm
{G})$, is defined as
$$
L(\mathrm
{G})=D(\mathrm
{G})-A(\mathrm
{G}),
$$
where $A(\mathrm{G})$ is the  $N\times N$ matrix given by $[A(\mathrm{G})]_{kj}=
1$  if $(j,k) \in \mathrm{E}$ and  $[A(\mathrm{G})]_{kj}=0$ otherwise, and $D(\mathrm
{G})={\rm diag}(d_1,\dots,d_N)$ with $d_k=\sum_{j=1,j\neq k}^N [A(\mathrm{G})]_{kj}$. By definition it is self-evident that zero is always an eigenvalue of $L(\mathrm
{G})$ for any directed graph $\mathrm
{G}$.

For any two digraphs sharing the same node set $\mathrm{G}_1=(\mathrm{V},\mathrm{E}_1)$ and $\mathrm{G}_2=(\mathrm{V},\mathrm{E}_2)$, we define their union as $\mathrm{G}_1\cup\mathrm{G}_2=(\mathrm{V},\mathrm{E}_1\cup \mathrm{E}_2)$.



\subsection{Linear algebra}
Given a matrix $M\in \mathbb{C}^{m\times n}$, the vectorization of $M$, denoted by ${\rm \bf vec}(M)$, is the $mn\times 1$ column vector  $([M]_{11}, \dots,  [M]_{m1},  [M]_{12}, \dots, [M]_{m2}, \dots, [M]_{1n},\dots, [M]_{mn})\T$. We have ${\rm \bf vec}(ABC)=(C\T\otimes A){\rm \bf vec}(B)$ for all  matrices $A,B,C$ with $ABC$ well defined, where $\otimes$ stands for the Kronecker product. We always use $I_\ell$ to denote the $\ell\times \ell$ identity matrix, and $\mathbf{1}_\ell$ for the all one vector in $\mathbb{R}^\ell$.

The following lemma is known as the Ger\v{s}gorin disc Theorem.

\medskip

\begin{lemma}\label{lemdisk} (pp. 344, \cite{Horn})
Let $A\in \mathbb{C}^{m\times m}$. Then all eigenvalues of $A$ are located in the union of $m$ discs
$$
\bigcup_{i=1}^m \Big\{z\in\mathbb{C}: \big|z-[A]_{ii}\big|\leq \sum_{j=1,j\neq i}^m \big|[A]_{ij}\big| \Big\}.
$$
\end{lemma}

A  matrix $A\in \mathbb{R}^{m\times m} $ is called a  nonnegative matrix if all its elements are nonnegative real numbers.  We call $A$ a {\it stochastic matrix} if $A \mathbf{1}_m =\mathbf{1}_m $, i.e., all the row sums of $A$ are equal to one. We call $A$ to be doubly stochastic if  both $A$ and $A\T$ are stochastic. A  matrix $A$ is called to be {\it irreducible} if $A$ cannot be conjugated into block upper triangular form by a permutation matrix $P$, i.e.,
\begin{align}
PAP^{-1}=
\left(\begin{array}{cc}
 E & F\\
 0 & G
 \end{array}\right),
\end{align}
where $E$ and $G$ are square matrices of sizes greater than zero.

For any nonnegative matrix $A\in\mathbb{R}^{N\times N} $, we can define its induced graph $\mathrm
{G}_A=(\mathrm {V}, \mathrm {E}_A)$ by that
$\mathrm{V}=\{1,\dots,N\}$ and $(i,j)\in \mathrm {E}_A $ if and only if $i\neq j$ and $[A]_{ji}>0$. A nonnegative matrix $A$ is irreducible if and only if $\mathrm
{G}_A$ is strongly connected. The following lemma is the famous Perron-Frobenius Theorem \cite{Perron-Frobenius} for irreducible nonnegative matrices.

\medskip

\begin{lemma}\label{lemmaperron-frobenius}
Let  $A$ be an irreducible nonnegative matrix. Then its spectral radius $\lambda(A)>0$ is a simple eigenvalue of $A$ which corresponds to a positive eigenvector.
\end{lemma}

\subsection{Quantum Mechanics}
\subsubsection{Quantum system and master equation}
 The state space associated with any isolated quantum system is a   Hilbert space which is a complex vector space with inner product \cite{Nielsen}. The state of a quantum system is  a unit vector in the system's state space.    For any Hilbert space $\mathcal{H}_\ast$,   it is convenient to use $|\cdot \rangle$, known as the Dirac notion, to denote  a unit (column)  vector in $\mathcal{H}_\ast$.  The complex conjugate of $|\xi \rangle$  is denoted as $\langle \xi|$. The state space of a composite quantum system is the tensor product, denoted~$\otimes$,  of the state space of each component system. For a quantum system, its state can also be described by a  density operator  $\rho$, which is Hermitian, positive in the sense that all its eigenvalues are non-negative, and   $\text{tr}(\rho)=1$.  For any $|p\rangle, |q\rangle \in \mathcal{H}_\ast$, we use the notion $|p\rangle \langle q|$ to denote the operator over $\mathcal{H}_\ast$ defined by
$$
\big(|p\rangle \langle q| \big) |\eta\rangle= \Big \langle  |q\rangle,  |\eta\rangle\Big\rangle |p\rangle, \ \ \forall |\eta\rangle \in \mathcal{H}_\ast,
$$
where $\Big\langle \cdot, \cdot \Big\rangle$ represents the inner product on  the Hilbert space   $\mathcal{H}_\ast$. In standard quantum mechanical notation the inner product $\Big \langle  |p\rangle,  |q\rangle\Big\rangle $ is denoted as $\big\langle p \big|q \big\rangle$.

 The evolution of the state $|\xi \rangle$ of a closed quantum system is described by the Schr\"{o}dinger equation. Equivalently these dynamics can be also written in forms of the evolution of the density operator $\rho$, called the  von Neumann equation. When a quantum system interacts with the environment,  a Markovian approximation can be applied
under the assumption of a short environmental correlation time permitting the neglect of memory
effects \cite{Breuer and Petruccione 2002}. Markovian master equations
have been widely used to model quantum systems with external inputs in quantum control, especially for Markovian quantum feedback \cite{Wiseman and Milburn 2009}. The so-called  Lindblad master equation is described as  \cite{Lindblad 1976}
\begin{equation}\label{LindbladMasterEquation}
\frac{d}{dt}{\rho}(t)=-\frac{\imath}{\hbar}[H,\rho(t)]+\sum_{k=1}^{K}\gamma_{k} \mathfrak{D}[L_{k}]\rho(t),
\end{equation}
where $H$ is  a Hermitian operator on the underlying Hilbert space known as the system Hamiltonian, $\imath^2=-1$, $\hbar$ is the reduced Planck constant, the non-negative coefficient $\gamma_{k}$'s
specify  the relevant relaxation rates, and $$\mathfrak{D}[L_{k}]\rho=L_{k}\rho L_{k}^{\dag}-\frac{1}{2}L_{k}^{\dag}L_{k}\rho-\frac{1}{2}\rho L_{k}^{\dag} L_{k}.$$ Here the $L_{k}$'s are the Lindblad operators representing the coupling of the system to the environment.

\subsubsection{Partial trace}
Let $\mathcal{H}_A$ and $\mathcal{H}_B$ be the state spaces of  two quantum systems $A$ and $B$, respectively. Their composite system is described as a density operator $\rho^{AB}$. Let $\mathfrak{L}_A$, $\mathfrak{L}_B$, and  $\mathfrak{L}_{AB}$ be the spaces of (linear) operators over  $\mathcal{H}_A$, $\mathcal{H}_B$, and  $\mathcal{H}_A\otimes\mathcal{H}_B$, respectively.   Then the partial trace over system $B$, denoted by ${\rm Tr}_{\mathcal{H}_B}$, is an operator mapping from $\mathfrak{L}_{AB}$ to $\mathfrak{L}_{A}$ defined by
$$
{\rm Tr}_{\mathcal{H}_B}\Big(|p_A\rangle  \langle q_A| \otimes  |p_B\rangle  \langle q_B| \Big)= |p_A\rangle  \langle q_A|  {\rm Tr} \Big(  |p_B\rangle  \langle q_B| \Big),\ \ \forall |p_A\rangle, |q_A \rangle\in \mathcal{H}_A, |p_B\rangle, |q_B\rangle \in \mathcal{H}_B.
$$

The reduced density operator (state) for system $A$, when the composite system is in the state  $\rho^{AB}$, is defined as $\rho^A= {\rm Tr}_{\mathcal{H}_B}(\rho^{AB})$. The physical interpretation of $\rho^A$ is that $\rho^A$ holds the full information of system $A$ in $\rho^{AB}$. For a detailed description  we hereby refer to \cite{Nielsen}.

\section{Problem Definition}\label{Secproblem}
In this section, we present the quantum network model and define the problem of interest.
\subsection{Qubit network and  permutation operators}
In quantum mechanical systems, a two-dimensional Hilbert space  forms  the most basic quantum system,  called a {\it qubit} system. Let $\mathcal{H}$ be a qubit space with a basis denoted  by $|0\rangle$ and $|1\rangle$. In this paper, we consider  a quantum network  with $n$ qubits indexed in the set $\mathsf{V}=\{1,\dots,n\}$  and  the state space of this $n$-qubit quantum network is denoted as the Hilbert space $\mathcal{H}^{\otimes n}=\mathcal{H}\otimes \dots \otimes \mathcal{H}$. The density operator of this $n$-qubit network is denoted as  $\rho$.

Interactions among the qubits are introduced by permutations. An $n$'th permutation is a bijection over $\mathsf{V}$, denoted by $\pi$. Denote the set of all $n$'th permutations as  $\mathfrak{P}$. There are $n!$ elements in $\mathfrak{P}$.  We can define the product of two permutations $\pi_1,\pi_2\in\mathfrak{P}$ as their composition, denoted $\pi_1  \pi_2\in \mathfrak{P}$,  in that
$$
\pi_1  \pi_2:\ \  \pi_1\pi_2(i)=\pi_1\big( \pi_2(i)\big),\ \ i\in\mathsf{V}.
$$
In this way the set  $\mathfrak{P}$ equipped with this product operation  defines a {\it group} known as the permutation group. Now associated with any $\pi\in \mathfrak{P}$, we define the corresponding  operator ${U}_\pi$ over $\mathcal{H}^{\otimes n}$, by
$$
{U_{\pi}} \big(q_{1}\otimes \dots \otimes q_n\big)= q_{\pi(1)}\otimes  \dots \otimes  q_{\pi(n)}, \ \ q_i\in \mathcal{H}, i=1,\dots,n.
$$
 In this way, the  operator $U_{\pi}$ permutes the states of the qubits. Particularly, a permutation $\pi$ is called a swapping between $j$ and $k$,   if  $\pi(j)=k$, $\pi(k)=j$, and $\pi(s)=s, s\in \mathsf{V}\setminus\{j,k\}$. It is straightforward to verify that $U_\pi$ is unitary, i.e., $U_\pi U_\pi^\dag=U_\pi^\dag U_\pi= I$.


The following definition provides a graphical  interpretation to  $U_{\pi}$.

\medskip

\begin{definition}
The quantum interaction graph associated with $U_{\pi}$, denoted $\mathsf{G}_\pi=(\mathsf{V}, \mathsf{E}_\pi)$, is the directed graph over $\mathsf{V}$ with
$
\mathsf{E}_\pi:=\big\{ \big(i,\pi(i)\big): i\neq \pi(i), i\in\mathsf{V}\big\}.
$
\end{definition}

\medskip

\begin{figure*}[t]
\begin{center}
\includegraphics[height=1.0in]{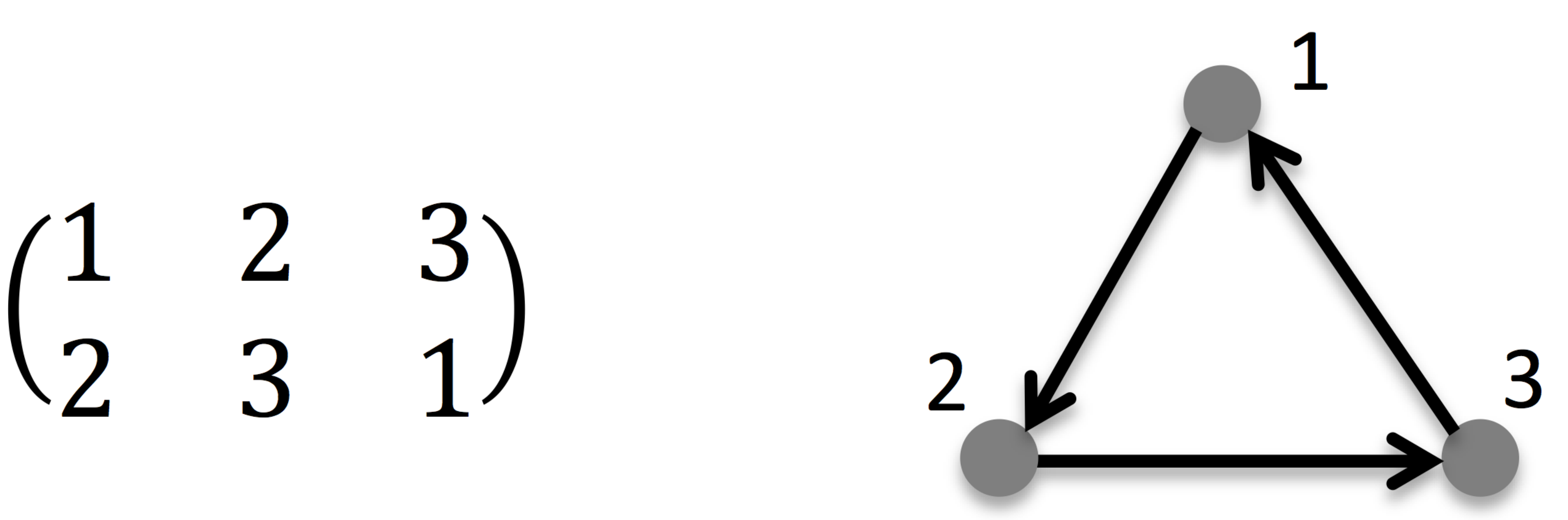}
\caption{The quantum interaction graph $\mathsf{G}_{\pi}$ over a three-qubit network for a given $\pi$ with $\pi(1)=2,\pi(2)=3,\pi(3)=1$.}
\label{cycle}
\end{center}
\end{figure*}

The  quantum interaction graph $\mathsf{G}_\pi$ indicates the information flow along the permutation operator $U_{\pi}$ (see Figure \ref{cycle} for an illustration).

\subsection{State evolution}
 Let $\mathfrak{P}_\ast \subseteq \mathfrak{P}$ be a subset of the permutation group.  In this paper, we are interested in  the state evolution of the quantum network  described by the following master equation
\begin{align}\label{sysconsensus}
\frac{d \rho}{dt}=\sum_{ \pi \in \mathfrak{P}_\ast}  \Big(U_{\pi}\rho U_{\pi}^\dag  -\rho\Big).
\end{align}

\begin{remark}
The system (\ref{sysconsensus})  was   proposed in \cite{Ticozzi-MTNS}  from a group theoretic perspective as well as  in \cite{ShiDongPetersenJohansson} when the permutation operators are restricted as swapping operators. If we choose a positive  number $w_{\pi}>0$ as the weight of  the permutation $\pi \in \mathfrak{P}_\ast$, the system (\ref{sysLind}) becomes
\begin{align}\label{weightedequation}
\frac{d \rho}{dt}=\sum_{ \pi \in \mathfrak{P}_\ast} w_{\pi} \Big(U_{\pi}\rho U_{\pi}^\dag  -\rho\Big).
\end{align}
Extension of the results established for (\ref{sysLind}) in the current paper to the weighted dynamics (\ref{weightedequation}), (even for the case with $w_{\pi}$ being time-dependent), is straightforward.
\end{remark}

\subsection{Objectives}

 Let $\mathcal{H}_i$ denote the two-dimensional Hilbert space corresponding to qubit $i$, $i\in\mathsf{V}$.   We denote by $$
\rho^k(t):= {\rm Tr}_{\otimes_{j\neq k} \mathcal{H}_j } \big(\rho(t)\big)
$$
 the reduced state of qubit $k$  at time $t$ for each $k=1,\dots,n$, where $\otimes_{j\neq k} \mathcal{H}_j$ stands for the remaining  $n-1$ qubits' space $\otimes_{j\neq k} \mathcal{H}_j$ and ${\rm Tr}_{\otimes_{j\neq k} \mathcal{H}_j }$ is the partial trace. Note that $\rho^k(t)$ contains all the information that qubit $k$ holds in the composite state $\rho(t)$.   Let $\mathfrak{C}_{\mathfrak{P}_\ast }$ be the subgroup  generated by $\mathfrak{P}_\ast$. We introduce   the following definition on the consensus notions of the quantum network (cf., \cite{Ticozzi,ShiDongPetersenJohansson}).

\medskip

 \begin{definition}
(i) The system (\ref{sysconsensus}) achieves global reduced-state average  consensus if
 \begin{align}
 \lim_{t\rightarrow \infty} \rho^j(t) =\frac{1}{n} \sum_{m=1}^n  \rho^m(0)
 \end{align}
 for all $ j\in\mathsf{V}$ and  initial states $\rho_0=\rho(0)$.

 \noindent (ii)  The system (\ref{sysconsensus}) achieves global $\mathfrak{P}_\ast$-average  consensus if
 \begin{align}
 \lim_{t\rightarrow \infty} \rho(t)=\frac{1}{\big|\mathfrak{C}_{\mathfrak{P}_\ast }\big|}\sum_{\pi\in\mathfrak{C}_{\mathfrak{P}_\ast } } U_\pi\rho_0  U_\pi^\dag,
 \end{align}
 for any initial state $\rho_0=\rho(0)$.

 \end{definition}

\medskip

 We term
$
\rho_{_{\mathfrak{P}_\ast}}:=\sum_{\pi\in\mathfrak{C}_{\mathfrak{P}_\ast } } U_\pi\rho  U_\pi^\dag/\big|\mathfrak{C}_{\mathfrak{P}_\ast }\big|
$
as the $\mathfrak{P}_\ast$-average of a given density operator $\rho \in \mathfrak{L}_{\mathcal{H}^{\otimes n}}$. The $\mathfrak{P}_\ast$-average is the arithmetic mean of the images of the permutation operators $U_\pi$ over $\rho$ for $\pi$ in the generated subgroup  $\mathfrak{C}_{\mathfrak{P}_\ast }$. When $\mathfrak{C}_{\mathfrak{P}_\ast }=\mathfrak{P}$, the $\mathfrak{P}_\ast$-average of $\rho$ becomes
$
\rho_{_{\mathfrak{P}}}=\sum_{\pi\in{\mathfrak{P}} } U_\pi\rho U_\pi^\dag/{n!},
$
which corresponds to the quantum {\it symmetric-state} consensus  introduced in \cite{Ticozzi}. As has been shown in \cite{Ticozzi}, for any $\rho$, its $\mathfrak{P}$-average $\rho_{_{\mathfrak{P}}}$ is  symmetric in the sense that it is invariant under any permutation operation, which  immediately implies that the  reduced states at each qubit are identical in $\rho_{_{\mathfrak{P}}}$.

\section{Unconditional Convergence from Perron-Frobenius Theory}\label{SecConvergence}

In this section, we make use of the  Perron-Frobenius theory to prove an unconditional convergence result for the system (\ref{sysconsensus}). The convergence speed is also explicitly characterized.

Recall that $\rho_{_{\mathfrak{P}_\ast}}$ be the $\mathfrak{P}_\ast$-average of  $\rho \in \mathfrak{L}_{\mathcal{H}^{\otimes n}}$.  Denote $\rho^k={\rm Tr}_{\otimes_{j\neq k} \mathcal{H}_j }(\rho)$. First of all, the following theorem  provides a tight criterion regarding when a $\mathfrak{P}_\ast$-average generates  identical reduced states.

 \medskip
\begin{theorem}\label{propositionSS}
(i) There holds  $\rho^i_{_{\mathfrak{P}_\ast}}=\rho^j_{_{\mathfrak{P}_\ast}}, \ i,j\in\mathsf{V}$ for all $\rho \in \mathfrak{L}_{\mathcal{H}^{\otimes n}}$ if and only if $ \mathsf{G}_{\mathfrak{P}_\ast}:=\bigcup_{\pi \in {\mathfrak{P}_\ast }} \mathsf{G}_\pi$ is strongly connected.

(ii) If $ \mathsf{G}_{\mathfrak{P}_\ast}$ is strongly connected, then  $
 \rho^k_{_{\mathfrak{P}_\ast}}= \rho^k_{_{\mathfrak{P}}}=\frac{1}{n} \sum_{m=1}^n  \rho^m
 $
 for all $k\in\mathsf{V}$.
\end{theorem}

\medskip

Next, the following theorem characterizes the convergence conditions for the system (\ref{sysconsensus}).

\medskip

\begin{theorem}\label{propositionconsensus}
The system (\ref{sysconsensus})  achieves global $\mathfrak{P}_\ast$-average  consensus.
\end{theorem}

\medskip

Note that Theorem  \ref{propositionconsensus} indicates that for the system  (\ref{sysconsensus}), the network's density operator converges to the   $\mathfrak{P}_\ast$-average of the initial density operator $\rho_0$. This convergence holds true for any choice of  $\mathfrak{P}_\ast$.  From the proof of Theorem \ref{propositionconsensus} we can even show that the convergence is exponential, with the exact convergence rate given by $$
\min_{\lambda_i\neq 0}{\rm Re}\big(\lambda_i(\mathrm{L}_\ast)\big),
$$
where
$
\mathrm{L}_\ast:=\sum_{\pi \in \mathfrak{P}_\ast}\Big(\mathrm{I}_{2^n}\otimes \mathrm{I}_{2^{n}}- \mathrm{U}_\pi\otimes \mathrm{U}_\pi\Big)
$
with $\mathrm{U}_\pi$ being the matrix representation of $U_\pi$ and $\otimes$ representing the Kronecker product, and $\lambda_i(\mathrm{L}_\ast)$ standing for an eigenvalue.

\begin{remark}
Theorems \ref{propositionSS} and  \ref{propositionconsensus} improve the previous results established in \cite{Ticozzi,Ticozzi-MTNS} in the following aspects: \begin{itemize}
 \item[(i)] It was shown in \cite{Ticozzi} that   $\mathfrak{P}$-average consensus implies reduced-state consensus, while Theorem \ref{propositionSS} further clarifies that $\mathfrak{P}_\ast$-average consensus can also imply  reduced-state consensus with a  necessary and sufficient condition given by  $ \mathsf{G}_{\mathfrak{P}_\ast}$ being strongly connected.
     \item[(ii)] It was proved in \cite{Ticozzi-MTNS}  that symmetric-state consensus is achieved if and only if $\mathfrak{P}_\ast$ is a generating subset of the permutation group, while Theorem   \ref{propositionconsensus} shows that the convergence statement is indeed unconditional with respect to the  choice of $\mathfrak{P}_\ast$.
         \end{itemize}
\end{remark}
 \subsection{Proof of Theorem \ref{propositionSS}}
The proof relies on some technical lemmas, whose proofs have been put in Appendix A.2, A.3, A.4, respectively.

%

\medskip

\begin{lemma}\label{lemcycle}
For any $\pi\in \mathfrak{P}$,  $\mathsf{G}_\pi$ is a  union of some   disjoint directed cycles.
\end{lemma}

%

\begin{lemma}\label{lemmaoperator}
For any $\pi \in \mathfrak{P}$ and $A_i\in \mathfrak{L}_{\mathcal{H}}$, we have $U_\pi\big( A_1\otimes \cdots\otimes A_n\big)U_\pi^\dag=  A_{\pi(1)}\otimes \cdots \otimes A_{\pi(n)} $.
\end{lemma}

\begin{lemma}\label{lemmastrongconnectivity}
The digraph $\bigcup_{\pi \in {\mathfrak{P}_\ast }} \mathsf{G}_\pi$ is strongly connected if and only if  $\bigcup_{\pi \in {\mathfrak{C}_{\mathfrak{P}_\ast} }} \mathsf{G}_\pi$ is fully connected.
\end{lemma}

\medskip

\noindent {\em Proof of (i). (Sufficiency.)} Take $\rho\in \mathfrak{L}_{\mathcal{H}^{\otimes n}}$. We denote by $\Theta_{\mathcal{H}}:=\big\{\theta_1,\dots, \theta_4\big\}$ a basis of $\mathfrak{L}_{\mathcal{H}}$ and write
$$
\rho=\sum_{i_1,\dots, i_n}C_{i_1 \dots i_n} \Big( \theta_{i_1}\otimes \cdots \otimes \theta_{i_n}\Big),
$$
where $C_{i_1 \dots i_n}\in \mathbb{C}$ and $\theta_{i_s}\in \Theta_{\mathcal{H}},s=1,\dots,n$. We now have
\begin{align}\label{17}
\rho^k_{_{\mathfrak{P}_\ast}} &={\rm Tr}_{\otimes_{j\neq k} \mathcal{H}_j }\bigg[\frac{1}{\big|\mathfrak{C}_{\mathfrak{P}_\ast }\big|} \sum_{\pi \in {\mathfrak{C}_{\mathfrak{P}_\ast} }} U_\pi \Big( \sum_{i_1,\dots, i_n}C_{i_1 \dots i_n} \big(\theta_{i_1}\otimes \cdots \otimes \theta_{i_n}\big) \Big) U_\pi^\dag \bigg]\nonumber\\
  &\stackrel{a)}{=} {\rm Tr}_{\otimes_{j\neq k} \mathcal{H}_j }\bigg[ \frac{1}{\big|\mathfrak{C}_{\mathfrak{P}_\ast }\big|} \sum_{\pi \in {\mathfrak{C}_{\mathfrak{P}_\ast} }}  \sum_{i_1,\dots, i_n}C_{i_1 \dots i_n} \Big( \theta_{i_{\pi(1)}}\otimes \cdots \otimes \theta_{i_\pi(n)} \Big) \bigg] \nonumber\\
    &\stackrel{b)}{=} \frac{1}{\big|\mathfrak{C}_{\mathfrak{P}_\ast }\big|} \sum_{\pi \in {\mathfrak{C}_{\mathfrak{P}_\ast} }}  \sum_{i_1,\dots, i_n}C_{i_1 \dots i_n} \Big[\theta_{i_{\pi(k)}}\prod _{j\neq k} {\rm Tr}(\theta_{i_{\pi(j)}})\Big],
\end{align}
where $a)$ holds from Lemma \ref{lemmaoperator} and $b)$ follows from the definition of partial trace.

Take $m\neq k\in \mathsf{V}$. Since by assumption  $\bigcup_{\pi \in {{\mathfrak{P}_\ast} }} \mathsf{G}_\pi$ is strongly connected,  $\bigcup_{\pi \in {\mathfrak{C}_{\mathfrak{P}_\ast }}} \mathsf{G}_\pi$ is fully connected according to Lemma \ref{lemmastrongconnectivity}. As a result, there exists $\pi_\ast\in\mathfrak{C}_{\mathfrak{P}_\ast }$ such that $\pi_\ast(m)=k$. We thus conclude
\begin{align}
\rho^m_{_{\mathfrak{P}_\ast}} &= \frac{1}{\big|\mathfrak{C}_{\mathfrak{P}_\ast }\big|} \sum_{\pi \in {\mathfrak{C}_{\mathfrak{P}_\ast} }}  \sum_{i_1,\dots, i_n}C_{i_1 \dots i_n} \Big[\theta_{i_{\pi(m)}}\prod _{j\neq m} {\rm Tr}(\theta_{i_{\pi(j)}})\Big]\nonumber \\
    &\stackrel{c)}{=} \frac{1}{\big|\mathfrak{C}_{\mathfrak{P}_\ast }\big|} \sum_{\pi\pi_\ast \in {\mathfrak{C}_{\mathfrak{P}_\ast} }}  \sum_{i_1,\dots, i_n}C_{i_1 \dots i_n} \Big[\theta_{i_{\pi\pi_\ast(m)}}\prod _{j\neq m} {\rm Tr}(\theta_{i_{\pi\pi_\ast(j)}})\Big]\nonumber\\
        &\stackrel{d)}{=} \frac{1}{\big|\mathfrak{C}_{\mathfrak{P}_\ast }\big|} \sum_{\pi\pi_\ast \in {\mathfrak{C}_{\mathfrak{P}_\ast} }}  \sum_{i_1,\dots, i_n}C_{i_1 \dots i_n} \Big[\theta_{i_{\pi(k)}}\prod _{j\neq \pi_\ast(m)} {\rm Tr}(\theta_{i_{\pi(j)}})\Big]\nonumber\\
                &\stackrel{e)}{=} \frac{1}{\big|\mathfrak{C}_{\mathfrak{P}_\ast }\big|} \sum_{\pi \in {\mathfrak{C}_{\mathfrak{P}_\ast} }}  \sum_{i_1,\dots, i_n}C_{i_1 \dots i_n} \Big[\theta_{i_{\pi(k)}}\prod _{j\neq k} {\rm Tr}(\theta_{i_{\pi(j)}})\Big]\nonumber\\
                &=\rho^k_{_{\mathfrak{P}_\ast}},
\end{align}
where $c)$ follows from the fact that $\{\pi: \pi\pi_\ast \in {\mathfrak{C}_{\mathfrak{P}_\ast} }\}={\mathfrak{C}_{\mathfrak{P}_\ast} }$ for any $\pi_\ast \in {\mathfrak{C}_{\mathfrak{P}_\ast} }$ since ${\mathfrak{C}_{\mathfrak{P}_\ast} }$ is by itself a group; $d)$ is from the selection of $\pi_\ast$ which satisfies $\pi_\ast(m)=k$; $e)$ holds again from  $\{\pi: \pi\pi_\ast \in {\mathfrak{C}_{\mathfrak{P}_\ast} }\}={\mathfrak{C}_{\mathfrak{P}_\ast} }$. This proves the sufficiency part of the theorem.

\medskip

\noindent {\em Proof of (i). (Necessity.)} If $ \mathsf{G}_{\mathfrak{P}_\ast}:=\bigcup_{\pi \in {\mathfrak{P}_\ast }} \mathsf{G}_\pi$ is not strongly connected, then $ \mathsf{G}_{\mathfrak{P}_\ast}$ is also not weakly connected by Lemmas \ref{lemmabalance} and \ref{lemcycle}. This means that $\mathsf{V}$ can be divided into two disjoint subsets $\mathsf{V}_1$ and $\mathsf{V}_2$ such that   $\rho_{\mathfrak{P}_\ast}$ never mixes the information of $\rho^k$'s  in $\mathsf{V}_1$ and $\mathsf{V}_2$. We can easily construct examples of $\rho_0$ based on this understanding under which $\rho^i_{_{\mathfrak{P}_\ast}}\neq \rho^j_{_{\mathfrak{P}_\ast}}$ for $i\in \mathsf{V}_1$ and $j\in \mathsf{V}_2$. This concludes the proof of Theorem \ref{propositionSS}. \hfill$\square$

\noindent {\em Proof of (ii)}. By (\ref{17}) we have
\begin{align}\label{21}
\rho^k_{_{\mathfrak{P}_\ast}}
    &{=} \frac{1}{\big|\mathfrak{C}_{\mathfrak{P}_\ast }\big|} \sum_{\pi \in {\mathfrak{C}_{\mathfrak{P}_\ast} }}  \sum_{i_1,\dots, i_n}C_{i_1 \dots i_n} \Big[\theta_{i_{\pi(k)}}\prod _{j\neq k} {\rm Tr}(\theta_{i_{\pi(j)}})\Big] \nonumber\\
    &=  \sum_{i_1,\dots, i_n}C_{i_1 \dots i_n} \bigg[\frac{1}{\big|\mathfrak{C}_{\mathfrak{P}_\ast }\big|} \sum_{\pi \in {\mathfrak{C}_{\mathfrak{P}_\ast} }}  \theta_{i_{\pi(k)}}\prod _{j\neq k} {\rm Tr}(\theta_{i_{\pi(j)}})\bigg].
\end{align}
Furthermore, from Lemma \ref{lemmastrongconnectivity}  $\bigcup_{\pi \in {\mathfrak{C}_{\mathfrak{P}_\ast} }} \mathsf{G}_\pi$ is fully connected when $\bigcup_{\pi \in {\mathfrak{P}_\ast }} \mathsf{G}_\pi$ is strongly connected. As a result, introducing
$$
\mathcal{I}_k(m):=\Big\{ \pi\in \mathfrak{C}_{\mathfrak{P}_\ast }:\ \pi(k)=m\Big\},
$$
we have $\big| \mathcal{I}_k(l)\big| =\big|\mathcal{I}_k(m)\big|$ for all $m,l\in\mathsf{V}$. Consequently, we conclude that
\begin{align}\label{19}
\frac{1}{\big|\mathfrak{C}_{\mathfrak{P}_\ast }\big|} \sum_{\pi \in {\mathfrak{C}_{\mathfrak{P}_\ast} }}  \theta_{i_{\pi(k)}}\prod _{j\neq k} {\rm Tr}(\theta_{i_{\pi(j)}})=\frac{1}{n}\sum_{s=1}^n  \theta_{i_{s}}\prod _{j\neq s} {\rm Tr}(\theta_{i_j}),
\end{align}
which does not depend on the choice of $\mathfrak{P}_\ast$.

Finally, plugging in (\ref{21}) with (\ref{19}) we obtain \begin{align*}
\rho^k_{_{\mathfrak{P}_\ast}}&=\frac{1}{n} \sum_{s=1}^n \sum_{i_1,\dots, i_n} C_{i_1 \dots i_n} \bigg[  \theta_{i_{s}}\prod _{j\neq s} {\rm Tr}(\theta_{i_j})\bigg]\nonumber\\
&=\frac{1}{n} \sum_{m=1}^n  \rho^m
\end{align*}
for all $k\in\mathsf{V}$. The proof of Theorem \ref{propositionSS} is complete.  \hfill$\square$
\subsection{Proof of Theorem \ref{propositionconsensus}}
We need  a few preliminary lemmas. The following lemma characterizes the fixed points of the so-called complete positive map.  Similar conclusion was drawn in \cite{Lindblad1999} and was later adapted to the following statement in \cite{PRAinformation} (Lemma 5.2).

\medskip

\begin{lemma}\label{lemmass}
Let $\mathcal{H}$ be a Hilbert space and denote by $\mathfrak{L}_{\mathcal{H}}$ the space of linear operators over $\mathcal{H}$. Define $T:\mathfrak{L}_{\mathcal{H}} \mapsto \mathfrak{L}_{\mathcal{H}}$ by
$$
T(X)=\sum_{j=1}^{K} M_j^\dag X M_j, \ \ X\in\mathfrak{L}_{\mathcal{H}}
$$
where $M_j \in \mathfrak{L}_{\mathcal{H}}$ for all $j$, $\sum_{j=1}^K M_j^\dag M_j=\sum_{j=1}^K M_j M_j^\dag=I$. Then, for any given $X_0\in \mathfrak{L}_{\mathcal{H}}$, $\sum_{j=1}^{K} M_j^\dag X_0 M_j=X_0$ if and only if $X_0M_j=M_j X_0, j=1,\dots,K$.
\end{lemma}

Define $\mathcal{K}_{\mathfrak{P}_\ast}:\mathfrak{L}_{\mathcal{H}^{\otimes n}} \mapsto \mathfrak{L}_{\mathcal{H}^{\otimes n}}$ by $\mathcal{K}_{\mathfrak{P}_\ast}(\rho)=\sum_{ \pi \in \mathfrak{P}_\ast}  \big(U_{\pi}\rho U_{\pi}^\dag -\rho \big)$. The following two lemmas hold.

\medskip

\begin{lemma}\label{lemmanull}
${\rm Null}(\mathcal{K}_{\mathfrak{P}_\ast}):=\big\{\rho:\mathcal{K}_{\mathfrak{P}_\ast}(\rho)=0 \big\}=\Big\{\rho:\rho=\sum_{\pi\in\mathfrak{C}_{\mathfrak{P}_\ast } } U_\pi\rho U_\pi^\dag/\big|\mathfrak{C}_{\mathfrak{P}_\ast }\big|\Big\}$.
\end{lemma}
{\it Proof.} From Lemma \ref{lemmass} we know $U_\pi \rho_0 =\rho_0 U_\pi, \pi \in\mathfrak{P}_\ast$ if $\mathcal{K}_{\mathfrak{P}_\ast}(\rho_0)=0$. Obviously $U_\pi \rho_0 =\rho_0 U_\pi, \pi \in\mathfrak{P}_\ast$ implies $U_\pi \rho_0 =\rho_0 U_\pi, \pi \in\mathfrak{C}_{\mathfrak{P}_\ast }$, which in turn leads to $\rho_0\in \big\{\rho=\sum_{\pi\in\mathfrak{C}_{\mathfrak{P}_\ast } } U_\pi\rho U_\pi^\dag/\big|\mathfrak{C}_{\mathfrak{P}_\ast }\big|\big\}$.

On the other hand, if $\rho_0=\sum_{\pi\in\mathfrak{C}_{\mathfrak{P}_\ast } } U_\pi\rho_0 U_\pi^\dag/\big|\mathfrak{C}_{\mathfrak{P}_\ast }\big|$, then $U_\pi \rho_0 U_\pi^\dag =\rho_0, \pi \in\mathfrak{P}_\ast$. This leads to that $\rho_0\in \big\{\rho:\mathcal{K}_{\mathfrak{P}_\ast}(\rho)=0 \big\}$. The proof is now complete. \hfill$\square$

\medskip

\begin{lemma}\label{lemmamultiplicity}
The zero eigenvalue of $\mathcal{K}_{\mathfrak{P}_\ast}$'s algebraic multiplicity is equal to its geometric multiplicity.
\end{lemma}
{\it Proof.} We make use of the Perron-Frobenius theorem to prove the desired conclusion. Let $\mathrm{U}_\pi$ denote the matrix representation of $U_{\pi}$ under the following basis of $\mathcal{H}^{\otimes n}$:
$$
\mathrm{B}:=\Big\{ |p_1\cdots p_n\rangle: \ |p_i\rangle\in\{|0\rangle, |1\rangle\}\Big\}.
$$
From the definition of  $U_\pi$ it is clear that the following claim holds.

\medskip

\noindent {\it Claim (i).} $\mathrm{U}_\pi$ is a doubly stochastic matrix  for any $\pi \in \mathfrak{P}$.

 We further consider the standard computational  basis of $\mathfrak{L}_{\mathcal{H}^{\otimes n}}$:
$$
\mathfrak{B}:=\Big\{ |p_1\cdots p_n\rangle\langle q_1\cdots q_n|: \ |p_i\rangle,|q_i\rangle\in\{|0\rangle, |1\rangle\}\Big\}
$$
We make another claim.

\medskip

\noindent {\it Claim (ii).} $U_{\pi}\big( |p_1\cdots p_n\rangle\langle q_1\cdots q_n| \big)U_{\pi}^\dag= \big|p_{\pi(1)}\cdots p_{\pi(n)}\big\rangle\big\langle q_{\pi(1)}\cdots q_{\pi(n)}\big|$.

This claim can be easily proved by verifying the images of the  above two operators are the same for all  $|\xi \rangle \in \mathrm{B}$. As a result, there is an order of the elements in $\mathfrak{B}$ under which the matrix representation of $\mathcal{K}_{\mathfrak{P}_\ast}(\cdot)$  is $$
\mathrm{K}_\ast:=\sum_{\pi \in \mathfrak{P}_\ast}\Big( \mathrm{U}_\pi\otimes \mathrm{U}_\pi-\mathrm{I}_{2^n}\otimes \mathrm{I}_{2^{n}}\Big),
$$
where $\otimes$ stands for the Kronecker product. Making use of Claim (i) we know that
$$
\Big(\mathbf{1}_{2^n}\T\otimes  \mathbf{1}_{2^n}\T\Big) \mathrm{U}_\pi\otimes \mathrm{U}_\pi= \Big(\mathbf{1}_{2^n}\T \mathrm{U}_\pi \Big) \otimes\Big(\mathbf{1}_{2^n}\T \mathrm{U}_\pi\Big)=\mathbf{1}_{4^n}\T
$$
and
$$
 \mathrm{U}_\pi\otimes \mathrm{U}_\pi\Big(\mathbf{1}_{2^n}\otimes  \mathbf{1}_{2^n}\Big)= \Big(\mathrm{U}_\pi \mathbf{1}_{2^n}\Big) \otimes\Big( \mathrm{U}_\pi\mathbf{1}_{2^n}\Big)=\mathbf{1}_{4^n},
$$
i.e., each $\mathrm{U}_\pi\otimes \mathrm{U}_\pi$ is a doubly  stochastic matrix.

We now focus on the matrix
$
\mathrm{H}_\ast:=\sum_{\pi \in \mathfrak{P}_\ast} \mathrm{U}_\pi\otimes \mathrm{U}_\pi
$
 and its induced graph $\mathrm{G}_{\mathrm{H}_\ast}$. Since every $\mathrm{U}_\pi\otimes \mathrm{U}_\pi$ is doubly  stochastic, $\mathrm{G}_{\mathrm{H}_\ast}$ is a balanced digraph. This further implies that every weakly connected component of   $\mathrm{G}_{\mathrm{H}_\ast}$ is balanced, and thus strongly connected by Lemma \ref{lemmabalance}. In other words, there exists a permutation matrix  $\mathrm{P}_\ast\in\mathbb{R}^{4^n\times 4^n}$ such that
 $$
 \mathrm{P}_\ast \mathrm{H}_\ast \mathrm{P}_\ast^{-1}={\rm diag}\big( \mathrm{P}_1,\dots,  \mathrm{P}_{c_0}  \big),
 $$
where each $\mathrm{P}_{i}$ is the adjacency matrix of each weakly connected  component  and $c_0$ stands for the number of those weakly connected components of $\mathrm{G}_{\mathrm{H}_\ast}$. Consequently, each $\mathrm{P}_{i}$ is irreducible.

Finally, applying the Ger\v{s}gorin disc Theorem, i.e., Lemma \ref{lemdisk}, we conclude that $\lambda(\mathrm{P}_{i})\leq |\mathfrak{P}_\ast|$. We also know $|\mathfrak{P}_\ast|$ is an eigenvalue of $\mathrm{P}_{i}$ due to the  stochasticity of each $\mathrm{U}_\pi\otimes \mathrm{U}_\pi$. Imposing the Perron-Frobenius theorem, i.e., Lemma \ref{lemmaperron-frobenius}, we further conclude that $|\mathfrak{P}_\ast|$ is a simple eigenvalue of every $\mathrm{P}_{i}$. This immediately yields that  the zero eigenvalue of $\mathrm{K}_\ast$'s algebraic multiplicity is equal to its geometric multiplicity since $\mathrm{K}_\ast=\mathrm{H}_\ast-|\mathfrak{P}_\ast|\cdot \mathrm{I}_{4^n}  $. The proof is  complete. \hfill$\square$

\medskip

We are now in a position to prove Theorem  \ref{propositionconsensus}. Lemma \ref{lemmamultiplicity} ensures that  the zero eigenvalue of $\mathrm{K}_\ast$'s algebraic multiplicity is equal to its geometric multiplicity. Lemma \ref{lemdisk} ensures that all non-zero eigenvalues of  $\mathrm{K}_\ast$ have negative real parts. These two facts imply that, for the system (\ref{sysconsensus}), $\rho(t)$ converges to a limit, and the limit is a fixed point, say $\rho_\ast$, in   the null space of $\mathcal{K}_{\mathfrak{P}_\ast}$. From Lemma~\ref{lemmanull} we know that
 \begin{align}\label{1}
 \rho_\ast=\sum_{\pi\in\mathfrak{C}_{\mathfrak{P}_\ast } } U_\pi\rho_\ast U_\pi^\dag/\big|\mathfrak{C}_{\mathfrak{P}_\ast }\big|.
 \end{align}
 From Lemma \ref{lemmass} we also see that
\begin{align}\label{2}
 &\frac{d}{dt}\sum_{\pi\in\mathfrak{C}_{\mathfrak{P}_\ast } } U_\pi\rho(t) U_\pi^\dag/\big|\mathfrak{C}_{\mathfrak{P}_\ast }\big|\nonumber\\
 &= \sum_{\pi\in\mathfrak{C}_{\mathfrak{P}_\ast } }U_\pi \sum_{\pi\in \mathfrak{P}_\ast}  \Big(U_{\pi}{\rho}(t) U_{\pi}^\dag  -{\rho}\Big)U_\pi^\dag/\big|\mathfrak{C}_{\mathfrak{P}_\ast }\big| \nonumber\\
 &=  \sum_{\pi\in\mathfrak{C}_{\mathfrak{P}_\ast } }U_\pi{\rho}(t) U_{\pi}^\dag /\big|\mathfrak{C}_{\mathfrak{P}_\ast }\big| - \sum_{\pi\in\mathfrak{C}_{\mathfrak{P}_\ast } }U_\pi{\rho}(t) U_{\pi}^\dag /\big|\mathfrak{C}_{\mathfrak{P}_\ast }\big|\nonumber\\
 &\equiv 0,
\end{align}
where we have used the fact that $\mathfrak{C}_{\mathfrak{P}_\ast }$ is a subgroup so that $\pi \mathfrak{C}_{\mathfrak{P}_\ast } =\mathfrak{C}_{\mathfrak{P}_\ast } $ for any $\pi\in\mathfrak{C}_{\mathfrak{P}_\ast }$. Therefore, combining  (\ref{1}) and (\ref{2}) we know that
 \begin{align}
\rho_\ast=\sum_{\pi\in\mathfrak{C}_{\mathfrak{P}_\ast } } U_\pi\rho_\ast U_\pi^\dag/\big|\mathfrak{C}_{\mathfrak{P}_\ast }\big|=\sum_{\pi\in\mathfrak{C}_{\mathfrak{P}_\ast } } U_\pi\rho_0 U_\pi^\dag/\big|\mathfrak{C}_{\mathfrak{P}_\ast }\big|.
 \end{align}
 We have now completed the proof of Theorem \ref{propositionconsensus}.

\section{The Missing Symmetry: A Graphical Look }\label{SecMissing}
In this section, we take a further look at  the incomplete symmetry in  the ${\mathfrak{P}_\ast}$-average, as reflected by the zero-pattern of the difference between ${\mathfrak{P}_\ast}$-average and ${\mathfrak{P}}$-average. We do this by investigating the dynamics of every element of the density operator along the master equation, from a graphical point of view. We show that  this graphical approach  not only provides a full characterization of  the missing symmetry, but also naturally leads to a much deeper understanding of  the original quantum dynamics.

\subsection{The information-flow hierarchy}
The quantum interaction graph $\mathsf{G}_\pi=(\mathsf{V}, \mathsf{E}_\pi)$ provides a characterization of the  information flow among the qubit network under  $U_\pi$. The ``resolution" of this characterization is however considerably low since merely the directions of the information flow are indicated  in $\mathsf{G}_\pi$. In order to provide  some more accurate characterizations of the information flow under $U_\pi$, we introduce the following definition by identifying the elements in the basis of $\mathcal{H}^{\otimes n}$ and $\mathfrak{L}_{\mathcal{H}^{\otimes n}}$ as classical nodes.

\medskip

\begin{definition}Let $\pi\in\mathfrak{P}$ and denote $\mathbf{V}=\big\{ |p_1\cdots p_n\rangle:\  p_i\in\{0, 1\}\big\}$ and $\mathscr{V}=\big\{ |p_1\cdots p_n\rangle\langle q_1\cdots q_n|: \  p_i, q_i\in\{0, 1\}\big\}$. Associated with the permutation  $\pi$, we define
\begin{itemize}
\item[(i)] the state-space graph $\mathbf{G}_\pi=(\mathbf{V}, \mathbf{E}_\pi)$ so that $\mathbf{E}_\pi$ consists of all  non-self-loop arcs in  $$
\Big\{\big(|p_1\cdots p_n\rangle,|p_{\pi(1)}\cdots p_{\pi(n)}\rangle\big):\ p_i\in\{0, 1\}\Big\};
$$

\item[(ii)] the operator-space  graph $\mathscr{G}_\pi=(\mathscr{V}, \mathscr{E}_\pi)$  so that  $\mathscr{E}_\pi$ consists of all  non-self-loop arcs in $$
\bigg\{\Big(|p_1\cdots p_n\rangle\langle q_1\cdots q_n|,|p_{\pi(1)}\cdots p_{\pi(n)}\rangle \langle q_{\pi(1)}\cdots q_{\pi(n)}|\Big):\  p_i, q_i\in\{0, 1\}\bigg\}.
$$
\end{itemize}
\end{definition}

\medskip

\begin{figure*}[t]
\begin{center}
\includegraphics[height=3.0in]{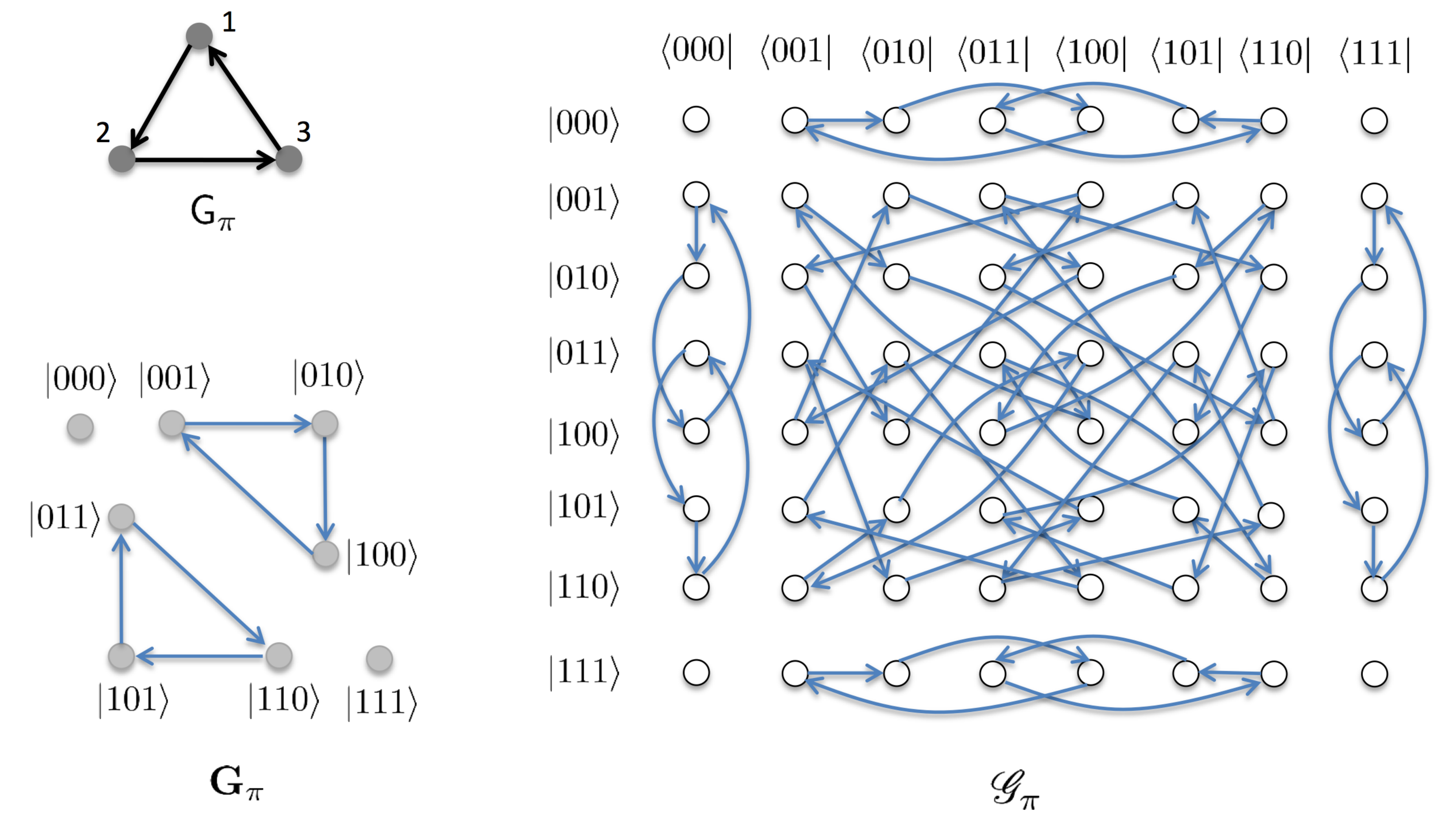}
\caption{The interaction graph $\mathsf{G}_\pi$, the state-space graph $\mathbf{G}_\pi$, and the operator-space graph $\mathscr{G}_\pi$. Various properties of these graphs can be verified from Theorems \ref{propstategraph} and \ref{propoperatorgraph}, e.g., $|\mathsf{E}_\pi|=6$, and $|\mathscr{E}_\pi|=2^4\times|\mathsf{E}_\pi|-|\mathsf{E}_\pi|^2=60$.}
\label{fig:threegraph}
\end{center}
\end{figure*}

From their definitions we see that both $\mathbf{G}_\pi$ and $\mathscr{G}_\pi$ are simple digraphs, and in fact they are always balanced from the nature of permutation indicated in Lemma \ref{lemcycle}. Note that $\mathbf{V}$ and $\mathscr{V}$ are basis of $\mathcal{H}^{\otimes n}$ and $\mathfrak{L}_{\mathcal{H}^{\otimes n}}$, respectively. Clearly $\mathbf{G}_\pi$ and $\mathscr{G}_\pi$ provide not only the directions of the information flow, but also the information itself in the flow of $U_\pi$. For the permutation $\pi(1)=2,\pi (2)=3,\pi(3)=1$ over a three-qubit network, its interaction graph $\mathsf{G}_\pi$, state-space graph $\mathbf{G}_\pi$, and operator-space graph $\mathscr{G}_\pi$, are respectively illustrated in Figure  \ref{fig:threegraph}.

Let $\mathbf{G}_{\mathfrak{P}}:=\bigcup_{\pi \in \mathfrak{P}} \mathbf{G}_{\pi}$. For  the state-space graph, we have  the following result.

\medskip

\begin{theorem}\label{propstategraph}
(i) Suppose $\mathsf{G}_\pi$ is a directed cycle. Then $\mathbf{G}_\pi$ has $2^n-2$ arcs for all $n\in \mathbb{N}^+$. Moreover, if $n\in \mathbb{N}^+$ is odd, then $\mathbf{G}_\pi$ has exactly $2+({2^n-2})/{n}$ strongly connected components, among which two are singletons and  the remaining  $({2^n-2})/{n}$ are directed cycles  with   size $n$.

(ii) $\mathbf{G}_{\mathfrak{P}}$ has $n+1$ strongly connected components with their sizes ranging from $n \choose 0$ to $n \choose n$, and each of its strongly connected components is fully connected. Consequently there are $$
E_n:={\sum_{k=1}^{n-1}} {n \choose k}\bigg[  {n \choose k} -1\bigg]
$$ arcs in  $\mathbf{G}_{\mathfrak{P}}$.
\end{theorem}
\medskip

\begin{remark}
From Lemma \ref{lemcycle}, $\mathsf{G}_\pi$ is a  union of  disjoint directed cycles for any $\pi$. This means that Theorem \ref{propstategraph}.(i) for $\pi$ whose interaction graph $\mathsf{G}_\pi$ is a directed cycle,   can be easily generalized to arbitrary permutations.
\end{remark}

For  the operator-space graph, the following result holds. Denote $\mathscr{G}_{\mathfrak{P}_\ast}:=\bigcup_{\pi \in \mathfrak{P}_\ast} \mathscr{G}_{\pi}$ and recall that $\mathcal{K}_{\mathfrak{P}_\ast}:\mathfrak{L}_{\mathcal{H}^{\otimes n}} \mapsto \mathfrak{L}_{\mathcal{H}^{\otimes n}}$ with $\mathcal{K}_{\mathfrak{P}_\ast}(\rho)=\sum_{ \pi \in \mathfrak{P}_\ast}  \big(U_{\pi}\rho U_{\pi}^\dag -\rho \big)$, and $\mathfrak{C}_{\mathfrak{P}_\ast}$ is the generated subgroup by $\mathfrak{P}_\ast$.

\medskip

\begin{theorem}\label{propoperatorgraph}
(i) $\big|\mathscr{E}_\pi \big|=2^{n+1}\big|\mathbf{E}_\pi\big|-\big|\mathbf{E}_\pi\big|^2$.

(ii) Suppose $\mathsf{G}_\pi$ is a directed cycle and $n\in \mathbb{N}^+$ is odd. Then $\mathscr{G}_\pi$ has exactly $4+({2^{2n}-4})/{n}$ strongly connected components, among which four are singletons and   the rest $({2^{2n}-4})/{n}$ are  directed cycles  with   size $n$.

(iii) There are a total of ${\rm dim}\big({\rm Null}(\mathcal{K}_{\mathfrak{P}_\ast})\big)$ strongly connected components in $\mathscr{G}_{\mathfrak{P}_\ast}$.

(iv) The components of $\mathscr{G}_{\mathfrak{C}_{\mathfrak{P}_\ast}}$ and $\mathscr{G}_{\mathfrak{P}_\ast}$ give the same partition of  $\mathscr{V}$, i.e., they agree on the same subsets of nodes.
\end{theorem}

\medskip

Theorems \ref{propstategraph} and \ref{propoperatorgraph} provide some detailed descriptions  of the information hierarchy for the quantum permutation operators, which can be quite useful in understanding the evolution of the quantum synchronization master equation. They are established via combinatorial  analysis  approach applied to  $\mathbf{E}_\pi$ and $\mathscr{E}_\pi$, whose detailed proofs are put in Appendix A.5 and A.6, respectively.

\begin{remark}
From Theorem  \ref{propoperatorgraph} and the proof of Theorem \ref{propositionconsensus}, we immediately know that whenever $n$ is odd,  the convergence rate under any permutation $\pi$ is equal to  the rate of convergence to a classical consensus over an $n$-node directed cycle. This rate can thus be explicitly given as
$$
1-\cos \Big( \frac{2\pi}{n}\Big),
$$
following the spectral  analysis to graphs (cf., Section 1.4.3, \cite{spectrabook}).
\end{remark}
\subsection{The zero pattern}
We now investigate the zero-pattern of the difference between ${\mathfrak{P}_\ast}$-average and ${\mathfrak{P}}$-average. Let
$$
\big[\rho_{_{\mathfrak{P}_\ast}}-\rho_{_{\mathfrak{P}}}\big]_{|p_1\cdots p_n\rangle\langle q_1\cdots q_n|}
$$
be the ${|p_1\cdots p_n\rangle\langle q_1\cdots q_n|}$-entry of $\rho_{_{\mathfrak{P}_\ast}}-\rho_{_{\mathfrak{P}}}$ under the basis $\mathscr{V}$. The following result holds.

\medskip

\begin{theorem}\label{thm3}
(i) There exists $\rho\in \mathfrak{L}_{\mathcal{H}^{\otimes n}}$ for which $\big[\rho_{_{\mathfrak{P}_\ast}}-\rho_{_{\mathfrak{P}}}\big]_{|p_1\cdots p_n\rangle\langle q_1\cdots q_n|}\neq 0$ if and only if the strongly connected components in $\mathscr{G}_{\mathfrak{P}}$ and $\mathscr{G}_{\mathfrak{P}_\ast}$ that contain  $|p_1\cdots p_n\rangle\langle q_1\cdots q_n|$, have different sizes.

(ii) Suppose  $\mathsf{G}_{\mathfrak{P}_\ast}$ is strongly connected. Then $\big[\rho_{_{\mathfrak{P}_\ast}}-\rho_{_{\mathfrak{P}}}\big]_{|p_1\cdots p_n\rangle\langle q_1\cdots q_n|}=0$ for all $\rho\in \mathfrak{L}_{\mathcal{H}^{\otimes n}}$ if one of the following conditions holds:

a) $p_1=\dots=p_n$ and $q_1=\dots=q_n$;

b) $p_1=\dots=p_n$, and $\sum_{i=1}^n q_i\in \{1,n-1\}$;

c) $|p_1\cdots p_n \rangle= |q_1\cdots q_n\rangle$, and $\sum_{i=1}^n q_i\in \{1,n-1\}$;

d) $|p_1\cdots p_n \rangle= |\bar{q}_1\cdots \bar{q}_n\rangle$, where $\bar{q}_i=1-q_i$, and $\sum_{i=1}^n q_i\in \{1,n-1\}$.
\end{theorem}
{\it Proof.} (i). The conclusion follows from the definition of $\mathscr{G}_{\mathfrak{P}_\ast}$.

(ii). We only need to make sure that the strongly components in $\mathscr{G}_{\mathfrak{P}}$ and $\mathscr{G}_{\mathfrak{P}_\ast}$ that contain  $|p_1\cdots p_n\rangle\langle q_1\cdots q_n|$ have the same size.

If $p_1=\dots=p_n$ and $q_1=\dots=q_n$, then $|p_1\cdots p_n\rangle\langle q_1\cdots q_n|$ is  an isolated node in $\mathscr{G}_{\mathfrak{P}}$. Thus a) always ensures the above same-size condition, actually for arbitrary $\mathfrak{P}_\ast$.

Now we move to Condition b) and suppose $p_1=\dots=p_n$ with  $\sum_{i=1}^n q_i=1$. Without loss of generality we let $q_1=1$. Since $\mathsf{G}_{\mathfrak{P}_\ast}$ is strongly connected, for any $i_\ast\in \mathsf{V}$, there exist $\pi_1,\dots,\pi_k$ such that $\pi_k \dots \pi_1(1)=i_\ast$. This implies that the component containing $|p_1\cdots p_n\rangle\langle q_1\cdots q_n|$ in $\mathscr{G}_{\mathfrak{P}_\ast}$ has $n$ nodes. From the choice of $|p_1\cdots p_n\rangle\langle q_1\cdots q_n|$ it is straightforward to see that the  component containing $|p_1\cdots p_n\rangle\langle q_1\cdots q_n|$ in $\mathscr{G}_{\mathfrak{P}}$  also has $n$ nodes. We can thus invoke (i) to conclude that $\big[\rho_{_{\mathfrak{P}_\ast}}-\rho_{_{\mathfrak{P}}}\big]_{|p_1\cdots p_n\rangle\langle q_1\cdots q_n|}=0$ for all $\rho\in \mathfrak{L}_{\mathcal{H}^{\otimes n}}$. While the other case in Condition b) with $\sum_{i=1}^n q_i=n-1$ holds from a symmetric argument.

Conditions c) and d) ensure the same-size condition in (i), via a  similar analysis as we use to investigate Condition b). We thus omit their details. The proof is now complete. \hfill$\square$

Theorem \ref{thm3}.(i) is  a tight graphical characterization of the missing symmetry in the  ${\mathfrak{P}_\ast}$-average. Theorem \ref{thm3}.(ii) further explicitly shows some symmetry kept in the  ${\mathfrak{P}_\ast}$-average when only reduced-state consensus is guaranteed (e.g., $\mathsf{G}_{\mathfrak{P}_\ast}$ is strongly connected, cf., Theorem \ref{propositionSS}).

\section{Switching Interactions, Synchronization, and Examples}\label{SecDiscussion}
In this section, we make use of the previously established results to further investigate the state evolution of the quantum network in the presence of  switching interactions and network Hamiltonian, respectively. We also provide a few numerical examples illustrating the results.
\subsection{Switching permutations}
We now study  the quantum synchronization master equation (\ref{sysLind}) subject to switching of permutation operators. To this end, we introduce $\mathscr{P}_\ast$ as the set containing  all the subsets of $\mathfrak{P}_\ast$, and a piecewise constant switching signal $\mu(\cdot): \mathbb{R}_{\geq 0} \mapsto \mathscr{P}_\ast$. We use $\mathfrak{P}_{\mu(t)}$ to denote the set of permutations selected at time $t$. Consider the following dynamics
\begin{align}\label{sysconsensusswitching}
\frac{d \rho}{dt}=\sum_{ \pi \in \mathfrak{P}_{\mu(t)}}  \Big(U_{\pi}\rho U_{\pi}^\dag  -\rho\Big).
\end{align}
which is evidently a time-varying version of (\ref{sysconsensus}).

For the ease of presentation we assume that there is a constant $\mu_D>0$ as a lower bound between any two consecutive switching instants of $\mu(\cdot)$. We introduce the following definition (cf., \cite{Touri,JSAC,juliencut} for related concepts in consensus dynamics over classical networks).

\medskip

\begin{definition}
We call $\pi\in \mathfrak{P}_\ast$ a persistent permutation under $\mu(\cdot)$ if
$$
\int_{t=0}^\infty \mathbf{1}_{\{\pi \in \mathfrak{P}_{\mu(t)}\}} dt =\infty.
$$
We further  introduce  $
 \mathfrak{P}^{^{\rm p}}_\ast:\big\{\pi: \pi\ \mbox{ is a persistent permutation}\big\}
 $ as the set of  persistent permutations.
\end{definition}

\medskip

 The following result holds, whose proof   is based on  the relationship between quantum and classical consensus dynamics, and the results  on classical consensus for the so-called ``cut-balanced graphs" established recently in \cite{juliencut}.

\medskip

\begin{theorem}\label{propositionswitchingconsensus}
The system (\ref{sysconsensusswitching}) ensures global $\mathfrak{P}_\ast $-average consensus  in the sense that $\lim_{t\to \infty }\rho(t)= \rho^0_{_{\mathfrak{P}_\ast }}$ for all $t_0\geq 0$ and all $\rho^0=\rho(t_0)$ if and only if $\mathfrak{C}_{ \mathfrak{P}^{^{\rm p}}_\ast}=\mathfrak{C}_{\mathfrak{P}_\ast}$.
\end{theorem}
{\it Proof.} The system (\ref{sysconsensusswitching})  has the form:
\begin{align}\label{classicalswitching}
\frac{d}{dt}\big[\rho(t)\big]_{|p_1\cdots p_n\rangle\langle q_1\cdots q_n|}=\sum_{\pi\in  \mathfrak{P}_{\mu(t)} } \Big( \big[\rho(t)\big]_{|p_{\pi^{-1}(1)}\cdots p_{\pi^{-1}(n)}\rangle \langle q_{\pi^{-1}(1)}\cdots q_{\pi^{-1}(n)}|}-\big[\rho(t)\big]_{|p_1\cdots p_n\rangle\langle q_1\cdots q_n|}\Big)
\end{align}
under the basis  $\mathscr{V}$. Note that (\ref{classicalswitching}) admits a classical consensus dynamics over the node set $\mathscr{V}$ with time-varying node interaction structures, where  at time $t$ node $|p_1\cdots p_n\rangle\langle q_1\cdots q_n|\in\mathscr{V}$ is influenced by its in-neighbors in the set
$$
\mathcal{N}^{^-}_{|p_1\cdots p_n\rangle\langle q_1\cdots q_n|}\big({t}\big):=\Big\{ |p_{\pi^{-1}(1)}\cdots p_{\pi^{-1}(n)}\rangle \langle q_{\pi^{-1}(1)}\cdots q_{\pi^{-1}(n)}|:\ \pi\in\mathfrak{P}_{\mu(t)} \Big\}.
$$
Similarly, the node $|p_1\cdots p_n\rangle\langle q_1\cdots q_n|\in\mathscr{V}$ influences its out-neighbors in the set
$$
\mathcal{N}^{^+}_{|p_1\cdots p_n\rangle\langle q_1\cdots q_n|}\big({t}\big):=\Big\{ |p_{\pi(1)}\cdots p_{\pi(n)}\rangle \langle q_{\pi(1)}\cdots q_{\pi(n)}|:\ \pi\in\mathfrak{P}_{\mu(t)} \Big\}.
$$

Note that for any $\pi\in\mathfrak{P}$, we  know that $\mathscr{G}_\pi$ is balanced. This immediately leads to
$$
\Big|\mathcal{N}^{^+}_{|p_1\cdots p_n\rangle\langle q_1\cdots q_n|}\big({t}\big)\Big|=\Big|\mathcal{N}^{^-}_{|p_1\cdots p_n\rangle\langle q_1\cdots q_n|}\big({t}\big)\Big|.
$$
As a result, this guarantees that (\ref{classicalswitching}) defines a cut-balanced classical consensus process in the sense that
$\deg^+_t\big(\mathscr{S}\big)=\deg^-_t\big(\mathscr{S}\big)$ for any node set $\mathscr{S}\subseteq \mathscr{V}$ and for any $t\geq 0$, where by definition
\begin{align}
\deg^+_t\big(\mathscr{S}\big):=\Big|\big\{z\in \mathscr{V}\setminus \mathscr{S}:  \exists v\in \mathscr{S}\ \mbox{and} \ \pi\in\mathfrak{P}_{\mu(t)}\  {\rm s.t.}\  (v,z)\in \mathscr{E}_\pi \big\} \Big|
\end{align}
and
\begin{align}
\deg^-_t\big(\mathscr{S}\big):=\Big|\big\{z\in \mathscr{V}\setminus \mathscr{S}:  \exists v\in \mathscr{S}\ \mbox{and} \ \pi\in\mathfrak{P}_{\mu(t)}\  {\rm s.t.}\  (z,v)\in \mathscr{E}_\pi \big\} \Big|.
\end{align}

Finally, by Theorem \ref{propoperatorgraph}.(iii)-(iv), there are ${\rm dim}\big({\rm Null}(\mathcal{K}_{\mathfrak{P}_\ast})\big)$ strongly connected components in $\mathscr{G}_{\mathfrak{C}_{\mathfrak{P}_\ast}}$, and thus the nodes in those different components can never interact under the dynamics (\ref{classicalswitching}).
Further we notice that    convergence to a $\mathfrak{P}_\ast$-average  is equivalent to componentwise  convergence to a classical average consensus over the strongly connected components (cf., \cite{ShiDongPetersenJohansson}).  Thus, the desired result holds directly from Theorem 1 in \cite{juliencut}, and this concludes the proof. \hfill$\square$

%

\subsection{Quantum  synchronization}

Let $H$  be the (time-invariant) Hamiltonian of the $n$-qubit quantum network.   We  consider the following master equation (cf., \cite{ShiDongPetersenJohansson})
\begin{align}\label{sysLind}
\frac{d \rho}{dt}=-\frac{\imath}{\hbar}[H, \rho]+\sum_{ \pi \in \mathfrak{P}_\ast}  \Big(U_{\pi}\rho U_{\pi}^\dag  -\rho\Big).
\end{align}

Let $H_0$ be  a Hermitian operator over $\mathcal{H}$. Denote the direct sum $H_0^{\oplus n}=\sum_{i=1}^n I^{\otimes (i-1)}\otimes H_0 \otimes I^{\otimes (n-i)}$. For the cases with $H= H_0^{ \oplus n}$ and $H= H_0^{ \otimes n}$, the results regarding the synchronization condition for the system  (\ref{sysLind}) are as follows, respectively.

\medskip

\begin{theorem}\label{thm1}
 Suppose $H= H_0^{ \oplus n}$. Then if and only if
 $
 \mathsf{G}_{\mathfrak{P}_\ast}
 $
is strongly connected, the system (\ref{sysLind}) achieves global  reduced-state synchronization in the sense that
\begin{align*}
\lim_{t\to \infty }\bigg\|\rho^k(t)-e^{-\imath H_0 t/\hbar} \bigg[{\rm Tr}_{\otimes_{j=1}^{n-1} \mathcal{H}_j }\Big( \frac{1}{\big|\mathfrak{C}_{\mathfrak{P}_\ast }\big|} \sum_{\pi\in\mathfrak{C}_{\mathfrak{P}_\ast } } U_\pi\rho_0 U_\pi^\dag\Big) \bigg]e^{\imath H_0 t/\hbar} \bigg\| =0
\end{align*}
for all $k\in \mathsf{V}$.
\end{theorem}

\medskip

\begin{theorem}\label{thm2}
 Suppose $H= H_0^{ \otimes n}$. Then if and in general only if $
 \mathsf{G}_{\mathfrak{P}_\ast}$ is strongly connected, the  system (\ref{sysLind}) achieves global reduced-state synchronization in this case that
 \begin{align}
\lim_{t\to \infty }\bigg\|\rho(t)-e^{-\imath H t/\hbar} \Big( \frac{1}{\big|\mathfrak{C}_{\mathfrak{P}_\ast }\big|} \sum_{\pi\in\mathfrak{C}_{\mathfrak{P}_\ast } } U_\pi\rho_0 U_\pi^\dag\Big)  e^{\imath H t/\hbar} \bigg\| =0.
\end{align}
  \end{theorem}
\medskip

Here by saying ``in general only if" in Theorem \ref{thm2}, we mean that we can always construct  examples of $H_0$ and qubit networks, under which strong connectivity of  $
 \mathsf{G}_{\mathfrak{P}_\ast}$ becomes essentially  necessary  for reduced-state synchronization. The proofs of the Theorems \ref{thm1} and \ref{thm2} are similar with the proof of Theorem 6 in \cite{ShiDongPetersenJohansson}, and are therefore omitted.

\subsection{Examples} \label{subnumerical}

Consider three qubits indexed in the set $\mathsf{V}=\{1,2,3\}$. We take $\mathfrak{P}_\ast=\{\pi_\ast\}$ with $\pi_\ast(1)=2,\pi_\ast(2)=3,\pi_\ast(3)=1$ as shown in Figure \ref{cycle}.   The corresponding  $\mathsf{G}_{\pi_\ast}$ is a directed cycle which is obviously strongly connected. The initial network state is chosen to be
$$
\rho_0=|10+\rangle \langle 10+|
$$
with $|+\rangle=\frac{1}{\sqrt{2}}\big(|0\rangle+|1\rangle\big)$.
The network Hamiltonian is chosen to be $H=\sigma_z\oplus\sigma_z\oplus \sigma_z$ or $H=\sigma_z\otimes \sigma_z\otimes \sigma_z$, where
\begin{equation}
\sigma_z=\left(\begin{matrix}
1 & 0\\
0 & -1\\
\end{matrix}\right)
\end{equation}
is  one of the Pauli matrices.

\subsubsection{Synchronization in reduced states}
\begin{figure*}
\begin{minipage}[t]{0.5\linewidth}
\centering
\includegraphics[width=2.8in]{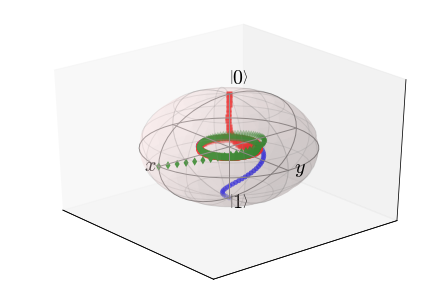}
\end{minipage}%
\begin{minipage}[t]{0.5\linewidth}
\centering
\includegraphics[width=2.8in]{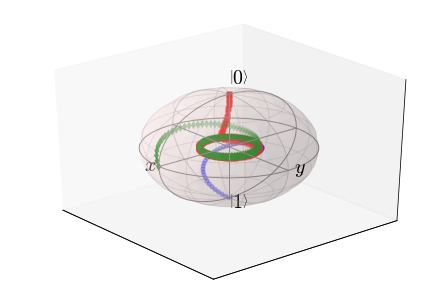}
\end{minipage}
\caption{The evolution of the reduced states of the three qubits for initial value $\rho_0=|10+\rangle \langle 10+|$ with  $H=\sigma_z\oplus\sigma_z\oplus \sigma_z$ (left), and  $H=\sigma_z\otimes\sigma_z\otimes \sigma_z$ (right), respectively.}
\label{fig:bloch}
\end{figure*}

We plot the evolution of the reduced states of the three qubits for the system (\ref{sysLind}) on one Bloch sphere with  initial value $\rho_0=|10+\rangle \langle 10+|$ for $H=\sigma_z\oplus\sigma_z\oplus \sigma_z$ and $H=\sigma_z\otimes\sigma_z\otimes \sigma_z$, respectively, in Figure \ref{fig:bloch}. The qubits' orbits  asymptotically tend to the same trajectory for both of the two cases. However due to the internal interactions raised by the tensor products in the network Hamiltonian, the evolution of the qubits' states gives different orbits for the two choices of $H$.

\begin{figure*}
\begin{minipage}[t]{0.5\linewidth}
\centering
\includegraphics[width=2.8in]{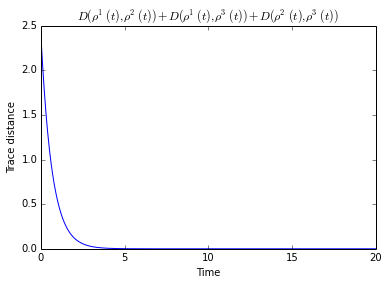}
\end{minipage}%
\begin{minipage}[t]{0.5\linewidth}
\centering
\includegraphics[width=2.8in]{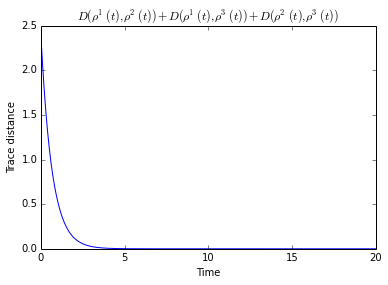}
\end{minipage}
\caption{The trace distance function $D(\rho^1(t),\rho^2(t))+D(\rho^1(t),\rho^3(t))+D(\rho^2(t),\rho^3(t))$ with  $H=\sigma_z\oplus\sigma_z\oplus \sigma_z$ (left), and  $H=\sigma_z\otimes\sigma_z\otimes \sigma_z$ (right), respectively.}
\label{fig:distance}
\end{figure*}

The trace distance between two density operator $\rho_1$, $\rho_2$ over the same Hilbert space, is defined as
$$
D(\rho_1,\rho_2)=\frac{1}{2}{\rm Tr}\sqrt{\Big(\rho_1-\rho_2\Big)^\dag\Big(\rho_1-\rho_2\Big)}.
$$
We plot the trace distance function $$
D(\rho^1(t),\rho^2(t))+D(\rho^1(t),\rho^3(t))+D(\rho^2(t),\rho^3(t))
$$ for the system (\ref{sysLind})  with  initial value $\rho_0=|10+\rangle \langle 10+|$, again for  $H=\sigma_z\oplus\sigma_z\oplus \sigma_z$, and  $H=\sigma_z\otimes\sigma_z\otimes \sigma_z$, respectively, in Figure \ref{fig:distance}.  Clearly they all converge to zero with an exponential rate and they show exactly the same convergence speed since the speed only depends on $\mathfrak{P}_\ast$, as discussed in the previous subsection.

\begin{figure*}
\begin{minipage}[t]{0.5\linewidth}
\centering
\includegraphics[width=2.8in, height=2.4in]{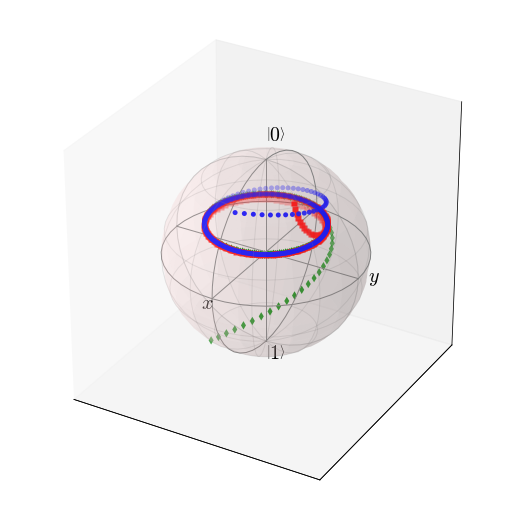}
\end{minipage}%
\begin{minipage}[t]{0.5\linewidth}
\centering
\includegraphics[width=2.8in, height=2.4in]{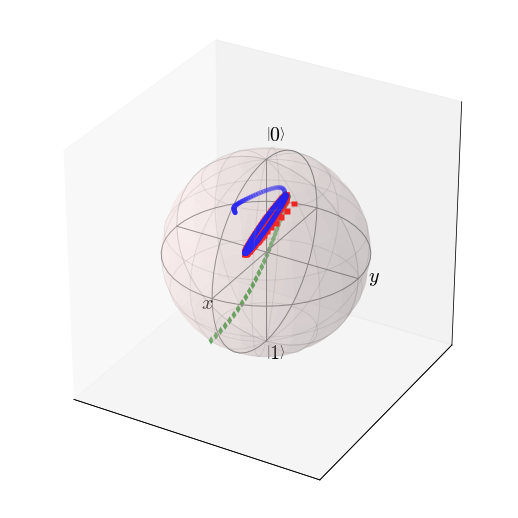}
\end{minipage}
\caption{The evolution of the reduced states of the three qubits for a different initial value with  $H=\sigma_z\oplus\sigma_z\oplus \sigma_z$ (left), and  $H=\sigma_z\otimes\sigma_z\otimes \sigma_z$ (right), respectively. Clearly when $H=\sigma_z\otimes\sigma_z\otimes \sigma_z$, drastic change appears for the shape of the limiting orbit compared to Figure \ref{fig:bloch}.}
\label{figure:different}
\end{figure*}

On the other hand, from Theorem \ref{thm1} we know that when  $H=\sigma_z\oplus\sigma_z\oplus \sigma_z$, the limiting orbit of  each qubit's reduced state is always parallel to the $x-y$ plane of the bloch sphere, no matter how the initial density operator is selected. In fact, we also know from Theorem \ref{thm1} that in this case the $z$-axis position of the limiting orbit is determined uniquely by the $\mathfrak{P}_\ast$-average of the initial network state. However,  when $H=\sigma_z\otimes\sigma_z\otimes \sigma_z$, there are internal interactions among the qubits, and as a result,  the shape of the limiting orbit under $H=\sigma_z\otimes\sigma_z\otimes \sigma_z$ is no longer predictable   with respect to the choice of  initial density operators. We illustrate this point in Figure \ref{figure:different}.

 \subsubsection{Partial symmetrization}
\begin{figure*}[t]
\begin{center}
\includegraphics[height=2.8in]{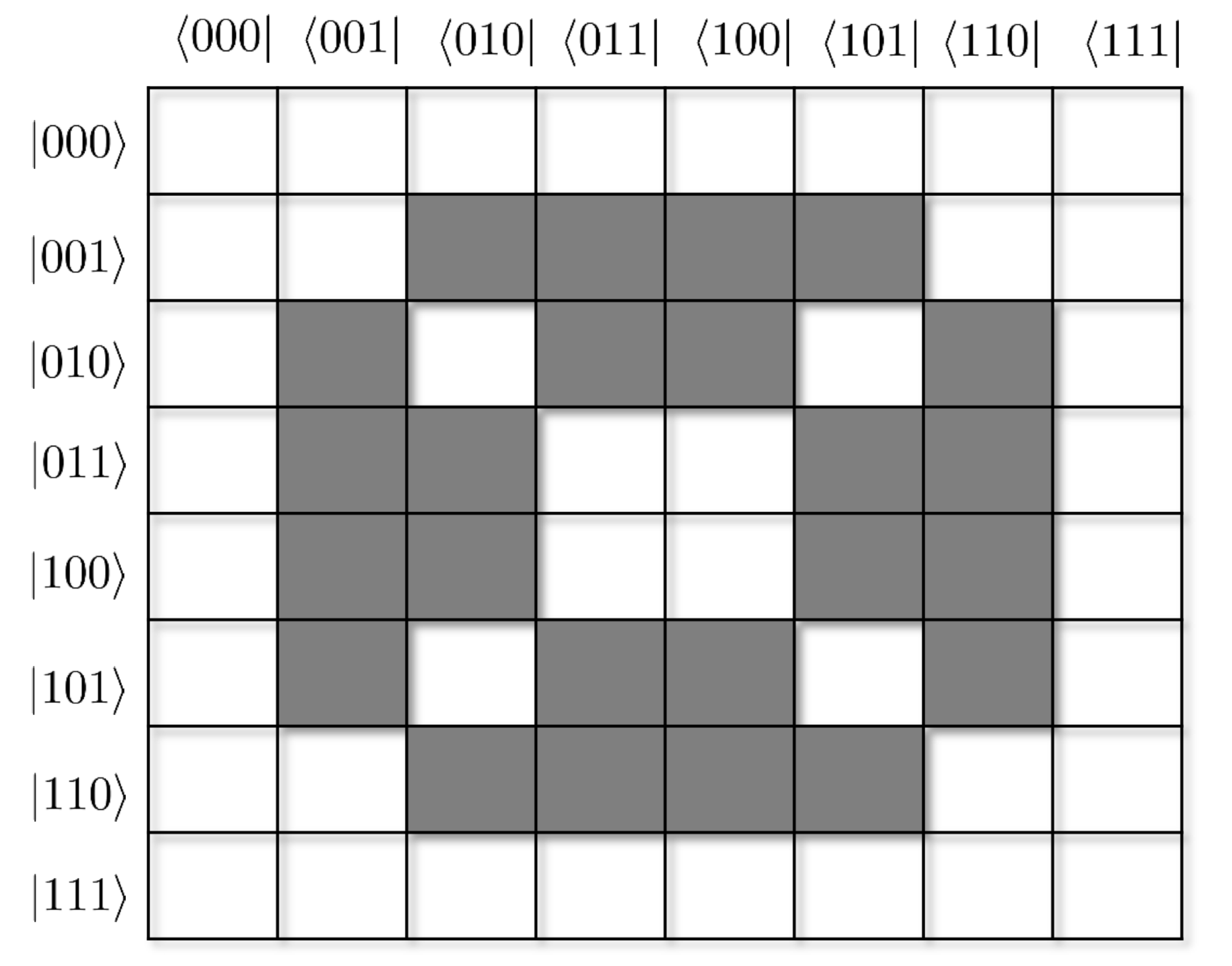}
\caption{The zero-pattern of the average difference $\rho_{_{\mathfrak{P}_\ast}}-\rho_{_{\mathfrak{P}}}$. Potential nonzero entries are shadowed.}
\label{fig:pattern}
\end{center}
\end{figure*}

We now investigate the difference between the $\mathfrak{P}_\ast$-average and the $\mathfrak{P}$-average
$$
\rho_{_{\mathfrak{P}_\ast}}-\rho_{_{\mathfrak{P}}}=\frac{1}{\big|\mathfrak{C}_{\mathfrak{P}_\ast }\big|}\sum_{\pi\in\mathfrak{C}_{\mathfrak{P}_\ast } } U_\pi\rho U_\pi^\dag-\frac{1}{n!}\sum_{\pi\in{\mathfrak{P}} } U_\pi\rho U_\pi^\dag.
$$
Under the standard computational  basis of $\mathfrak{L}_{\mathcal{H}^{\otimes 3}}$
$$
\Big\{ |p_1p_2p_3\rangle\langle q_1q_2 q_3|: \ |p_i\rangle,|q_i\rangle\in\{|0\rangle, |1\rangle\}\Big\}
$$
we plot the zero-pattern for the entries of $\rho_{_{\mathfrak{P}_\ast}}-\rho_{_{\mathfrak{P}}}$ with the given $\mathfrak{P}_\ast=\{\pi_\ast\}$ in Figure \ref{fig:pattern}. The zero-pattern of $\rho_{_{\mathfrak{P}_\ast}}-\rho_{_{\mathfrak{P}}}$ is obtained as follows: we randomly select  $\rho$, and shadow every  entry that  can be nonzero among the selections. From Figure \ref{fig:pattern} we clearly see the missing symmetry in $\rho_{_{\mathfrak{P}_\ast}}$ indicated by the zero-pattern, which by itself shows certain symmetry. The figure is consistent with the result in Theorem \ref{thm3}.

\section{Conclusions}\label{SecConclusion}
 This paper presented a systematic study of the consensus seeking  of quantum  networks  under directed interactions defined  by a set of permutation operators over the network whose state evolution is described by a continuous-time master equation. We established an unconditional convergence result indicating that the quantum network state always converges with the limit determined by the generating subgroup of the permutations making use of the Perron-Frobenius theory. A tight graphical criterion regarding when such limit admit a reduced-state consensus was also obtained.  Further, we provided a full characterization to the missing symmetry in the reduced-state consensus from a graphical point of view, where  the information-flow  hierarchy in quantum permutation operators is characterized by  different layers of information-induced graphs. Finally, we investigated   quantum synchronization conditions, characterized  quantum consensus under switching interactions applying the recent work of  Hendrickx and  Tsitsiklis. Numerical examples were also given illustrating the obtained results. Interesting future work includes potential decoherence in the quantum synchronization under general network Hamiltonian as well as more design of local quantum interactions that generate richer or more useful limiting states.

\section*{Acknowledgements}

The authors wish to acknowledge useful discussions with Matthew James and Luis Espinosa. The authors also thank Alain Sarlette for his kind suggestions in improving the presentation and clarity  of the results.  This work is supported in part  by the Australian Research Council under project FL110100020, the  Australian Research Council Centre of Excellence for Quantum Computation and Communication Technology (project number CE110001027), and AFOSR Grant FA2386-12-1-4075.

\medskip

\section*{Appendix}

\subsection*{A.1 Proof of Lemma \ref{lemmabalance}} If a digraph  $\mathrm
{G}=(\mathrm {V}, \mathrm {E})$ is balanced, then apparently it must hold that $\deg^+(S)=\deg^-(S)$ for any node set $S\subseteq \mathrm{V}$, where by definition
\begin{align}
\deg^+(S):=\Big|\big\{z\in \mathrm{V}\setminus S:  \exists v\in S\ {\rm s.t.}\ (v,z)\in \mathrm{E} \big\} \Big|
\end{align}
and
\begin{align}
\deg^-(S):=\Big|\big\{z\in \mathrm{V}\setminus S:  \exists v\in S\ {\rm s.t.}\ (z,v)\in \mathrm{E} \big\} \Big|.
\end{align}

Only the necessity statement needs to be verified. Suppose $\mathrm
{G}=(\mathrm {V}, \mathrm {E})$ is not strongly connected. Then there exists a partition of $\mathrm{V}$ into two nonempty and disjoint subsets of nodes $\mathrm {V}_1$ and $\mathrm{V}_2$ such that $\mathrm{V}_1\cup \mathrm{V}_2=\mathrm{V}$, for which there is no arc leaving from $\mathrm{V}_1$ pointing to $\mathrm{V}_2$. On the other hand the graph is weakly connected, so there exists at least one arc from $\mathrm{V}_2$  to $\mathrm{V}_1$. Taking $S=\mathrm{V}_1$ in the above argument we reach a contradiction. This concludes the proof.

\subsection*{A.2 Proof of Lemma \ref{lemcycle}}
 Take $\pi \in \mathfrak{P}$ and $i\in \mathsf{V}$. Since $\pi$ is a bijection over $\mathsf{V}$ there must exist an integer $K \geq 1$ such that $\pi^{K}(i)=i$. Without loss of generality we can assume such $K$ has been taken as  the smallest integer satisfying $\pi^{K}(i)=i$ and $K\geq 2$. Note that it is impossible that $\pi^{k_1}(i)=\pi^{k_2}(i)$ for some $0\leq k_1<k_2 \leq K$ since otherwise $\pi^{k_2-k_1}(i)=i$, which contradicts the choice of $K$. This means that
$$
i,\pi(i),\cdots, \pi^K(i)=i
$$
admits a directed cycle in  $\mathsf{G}_\pi$. Examining  every $i\in \mathsf{V}$ using the above argument concludes the lemma immediately.

\subsection*{A.3 Proof of Lemma \ref{lemmaoperator}}
From  the definition of $U_\pi$ we know that $U_\pi^\dag=U_\pi^{-1}=U_{\pi^{-1}}$, where $\pi^{-1}$ is the inverse of $\pi$ in the permutation group $\mathfrak{P}$.

The following equalities hold:
\begin{align}
&\Big \langle p_1\cdots p_n \Big| U_\pi \Big( A_1\otimes \cdots \otimes A_n \Big) U_\pi ^\dag \Big| q_1\cdots q_n\Big\rangle\nonumber\\
=&\Big \langle p_{\pi^{-1}(1)}\cdots p_{\pi^{-1}(n)} \Big|  \Big( A_1\otimes \cdots \otimes A_n \Big) \Big| q_{\pi^{-1}(1)}\cdots q_{\pi^{-1}(n)}\Big\rangle\nonumber\\
=& \sum_{i=1}^{n} \big \langle p_{\pi^{-1}(i)}\big|  A_i \big|  q_{\pi^{-i}(1)}\big\rangle\nonumber\\
=& \sum_{\pi(i)=1}^{n} \big \langle p_{i}\big|  A_{\pi(i)} \big|  q_{i}\big\rangle\nonumber\\
=& \sum_{i=1}^{n} \big \langle p_{i}\big|  A_{\pi(i)} \big|  q_{i}\big\rangle\nonumber\\
=&\Big \langle p_1\cdots p_n \Big|  A_{\pi(1)}\otimes \cdots \otimes A_{\pi(n)} \Big| q_1\cdots q_n\Big\rangle
\end{align}
for all $|p_i\rangle, |q_i\rangle \in \mathcal{H}, i\in\mathsf{V}$. This concludes the proof.

\subsection*{A.4 Proof of Lemma \ref{lemmastrongconnectivity}}
Apparently only the sufficiency statement needs to be proved.  From Lemma \ref{lemcycle} we know that for any $\pi \in \mathfrak{P}$, there is an integer $K \geq 1$ such that $\pi^{K}=I$. This means that $\pi^{-1}=\pi^{K-1}$.

From the definition of $\mathsf{G}_\pi$, $\bigcup_{\pi \in {\mathfrak{C}_{\mathfrak{P}_\ast} }} \mathsf{G}_\pi$ being fully connected is equivalent to that for any two nodes $i\neq j\in \mathsf{V}$, there exists a permutation $\pi_\ast\in\mathfrak{C}_{\mathfrak{P}_\ast}  $ such that
\begin{align}\label{3}
\pi_\ast (i)=j.
\end{align}
Note that the above argument yields that any $\pi_\ast\in\mathfrak{C}_{\mathfrak{P}_\ast}$ can be written as  $\pi_\ast=\pi_k \cdots \pi_1 $ with  $\pi_s\in \mathfrak{P}_\ast, 1\leq s\leq k$. In other words, (\ref{3}) leads to
\begin{align}\label{5}
\pi_{k} \dots \pi_{1} (i)=j, \ \ \pi_s\in \mathfrak{P}_\ast,\ 1\leq s\leq k.
\end{align}
 We immediately conclude from (\ref{5}) that $\bigcup_{\pi \in {{\mathfrak{P}_\ast} }} \mathsf{G}_\pi$ is strongly connected.

\subsection*{A.5 Proof of Theorem \ref{propstategraph}}
Recall that for a positive integer $n$, two integers $a$ and $b$ are said to be congruent modulo $n$, denoted
$
a=b \pmod n,\,
$
if $n$ divides their difference $a-b$.

\noindent (i). Since $\mathsf{G}_\pi$ is a directed cycle, without loss of generality we assume that $\pi(i)=i+1 \mod n$. Suppose
$$
|p_1\cdots p_n\rangle=|p_{\pi(1)}\cdots p_{\pi(n)}\rangle.
$$
Then we obtain $p_1=p_2=\dots =p_n$ from the definition of $\pi$. This implies that $\big(|p_1\cdots p_n\rangle,|p_{\pi(1)}\cdots p_{\pi(n)}\rangle\big)$
defines an arc in $\mathbf{E}_\pi$ as long as $p_1=p_2=\dots =p_n$ does not hold. We immediately conclude that $\big|\mathbf{E}_\pi\big|=2^n-2$.

Now we investigate the property of the strongly connected components of $\mathbf{G}_\pi$. Note that as $\mathbf{G}_\pi$ is apparently balanced, each of its weakly connected components is strongly connected by Lemma~\ref{lemmabalance}.  The following lemma holds.
\begin{lemma}\label{lemmamod}
Suppose $n\geq 3$ is an odd integer and $\mathsf{G}_\pi$ associated with $\pi\in \mathfrak{P}$  is a directed cycle. Then  $\mathsf{G}_{\pi^{k}}$ is also a directed cycle for all $k\neq 0 \pmod n$.
\end{lemma}
{\it Proof.} Again without loss of generality we assume that $\pi(i)=i+1 \mod n$.

 We first  prove the conclusion for $k=2$.  By Lemma \ref{lemcycle} we only need to show that  $\mathsf{G}_{\pi^2}$ is strongly connected.  Take $i_\ast\neq j_\ast\in\mathsf{V}$. The following  modular equation (with respect to $x$)
\begin{align}\label{11}
i_\ast+2x=j_\ast \pmod n
\end{align}
always has a solution since $n\geq 3$ is an odd integer. Let $x_0 \in \mathbb{N}$ be a solution of (\ref{11}).  Then
\begin{align*}
\big(\pi^2\big)^{x_0}(i_\ast)=j_\ast,
\end{align*}
which yields a path from $i_\ast$ to $j_\ast $ in $\mathsf{G}_{\pi^{2}}$ with length $x_0$. This proves that $\mathsf{G}_{\pi^2}$ is strongly connected, which must be a directed cycle.

Now let $0\leq k \leq n-1$. Since $\pi^n=I$, for any $k\neq 0\ ({\rm mod}\ n)$ we can find a positive integer $\gamma$ satisfying $\pi^k=\big(\pi^{\gamma}\big)^2$. As a result,  the overall conclusion  follows from  a straightforward induction argument. This completes the proof. \hfill$\square$

\begin{lemma}\label{lemmacomponentstategraph}
Suppose  $\mathsf{G}_\pi$ is a directed cycle and let $n$ be an odd integer. Take $|p_1\cdots p_n\rangle$ with $p_i\in \{0,1\}$ for all $i$ and assume that  at least two $p_i$'s take  distinct values. Then the elements in
$$
\Big\{|p_{\pi^k(1)}\cdots p_{\pi^k(n)}\rangle, k=0,1,\dots, n-1\Big\}
$$
are pairwise distinct.
\end{lemma}
{\it Proof.} Suppose there are $0\leq k_\ast< l_\ast\leq n-1$ such that $|p_{\pi^{k_\ast}(1)}\cdots p_{\pi^{k_\ast}(n)}\rangle=|p_{\pi^{l_\ast}(1)}\cdots p_{\pi^{l_\ast}(n)}\rangle$. This immediately gives $|p_{1}\cdots p_{n}\rangle=|p_{\tilde{\pi}(1)}\cdots p_{\tilde{\pi}(n)}\rangle$, where $\tilde{\pi}:=\pi^{l_\ast-k_\ast}$.

   Note that $\tilde{\pi}$ must be a permutation whose interaction graph $\mathsf{G}_{\tilde{\pi}}$ is a directed cycle from Lemma \ref{lemmamod}. Consequently, for any $i \neq j \in \mathsf{V}$, there is a path from $i$ to $j$ in  $\mathsf{G}_{\tilde{\pi}}$. In other words,  there exists an positive integer $z_0$ such that $j=\tilde{\pi}^{z_0}(i)$. This implies $p_i=p_j$ observing the equality $|p_{1}\cdots p_{n}\rangle=|p_{\tilde{\pi}(1)}\cdots p_{\tilde{\pi}(n)}\rangle$. Since $i$ and $j$ are chosen arbitrarily, we conclude
 $p_1=p_2=\cdots=p_n$, which  contradicts our standing assumption. We have now proved  the lemma. \hfill$\square$

From Lemma \ref{lemmacomponentstategraph} and noticing $\pi^n=I$, we immediately conclude that for any $|p_1\cdots p_n\rangle$ with at least two $p_i$'s taking  distinct values,
$$
\Big\{|p_{\pi^k(1)}\cdots p_{\pi^k(n)}\rangle, k=0,1,\dots, n-1\Big\}
$$
defines the  set of nodes  to which there is a path  from $|p_1\cdots p_n\rangle$ in $\mathbf{G}_\pi$. Consequently, the component where $|p_1\cdots p_n\rangle$ locates contains  exactly $n=\Big| \big\{|p_{\pi^k(1)}\cdots p_{\pi^k(n)}\rangle, k=0,1,\dots, n-1\big\}\Big|$ nodes and $n$ directed arcs. Invoking the fact that there are a total of $2^n-2$ arcs in $\mathbf{G}_\pi$, such components with a size $n$ count $(2^n-2)/n$. The total number of components are certainly $(2^n-2)/n+2$ (two singleton components corresponds to $p_1=\cdots=p_n=0$ and $p_1=\cdots=p_n=1$, respectively). The fact that each non-singleton component is a directed cycle simply  follows from  that $\pi^n=I$.

\medskip
\noindent (ii) Suppose $|p_1\cdots p_n\rangle, |q_1\cdots q_n\rangle\in \mathbf{V}$ satisfy $\sum_{i=1}^n p_i =\sum_{i=1}^n q_i =k$ with $0\leq k\leq n$. Then obviously we can find a $\pi\in \mathfrak{P}$ such that $|q_1\cdots q_n\rangle=|p_{\pi(1)}\cdots p_{\pi(n)}\rangle$. This immediately leads to that the subset of nodes
$$
\Big\{ |p_1\cdots p_n\rangle,\ \sum_{i=1}^n p_i=k\Big\}
$$
induces a fully connected component of $\mathbf{G}_\mathfrak{P}$. The rest of the conclusions follows from direct computations.

The proof of Theorem \ref{propstategraph} is now complete.

\subsection*{A.6 Proof of Theorem \ref{propoperatorgraph}}
(i). Let $\mathrm{U}_\pi$ be the matrix representation of the operator $U_\pi$ under the basis  $\mathbf{V}$ of $\mathcal{H}^{\otimes n}$.  From the definition of $U_\pi$ and $\mathbf{G}_\pi$ we see that $\mathrm{U}_\pi$ is exactly the  adjacency matrix of $\mathbf{G}_\pi$. Define $\mathcal{J}_\pi: \mathfrak{L}_{\mathcal{H}^{\otimes n}} \mapsto \mathfrak{L}_{\mathcal{H}^{\otimes n}}$ by that
$$
\mathcal{J}_\pi(\rho)= U_\pi \rho U_\pi^\dag,\ \ \rho \in \mathfrak{L}_{\mathcal{H}^{\otimes n}}.
$$
From the correspondence of tensor product and Kronecker product we see that $\mathrm{U}_\pi \otimes \mathrm{U}_\pi$ is a matrix representation of $\mathcal{J}_\pi$ under the basis $\mathscr{V}$, as well as the adjacency matrix of $\mathscr{G}_\pi$.

Suppose $\big| \mathbf{E}_\pi\big|=m$. There is a permutation matrix $\mathrm{P}\in\mathbb{R}^{2^n \times 2^n}$ such that
\begin{align}\label{9}
\mathrm{P} \mathrm{U}_\pi \mathrm{P}^{-1}= \tilde{\mathrm{U}}_\pi=
\left(\begin{array}{cc}
 \mathrm{I}_{2^n-m} & 0\\
 0 & \mathrm{Q}_\pi
 \end{array}\right)
\end{align}
with $\mathrm{Q}_\pi$ being a stochastic matrix with zero diagonals. It is therefore straightforward to directly compute that there are
$$
m^2 + 2m(2^n-m)= 2^{n+1}m-m^2
$$
nonzero and non-diagonal entries in $\mathrm{U}_\pi\otimes \mathrm{U}_\pi$, which immediately yields the desired conclusion.

\medskip

\noindent (ii). The conclusion follows immediately combining Theorem \ref{propstategraph}.(i) and the structure of $\tilde{\mathrm{U}}_\pi$ shown in (\ref{9}).

\medskip

\noindent (iii). By definition
$$
{L}(\mathscr{G}_{\mathfrak{P}_\ast}):=\sum_{\pi\in \mathfrak{P}_\ast} \Big(\mathrm{I}_{4^n}-\mathrm{U}_\pi\otimes \mathrm{U}_\pi\Big)
$$
is the Laplacian of $\mathscr{G}_{\mathfrak{P}_\ast}$. Recall  that every weakly strongly connected component of $\mathscr{G}_{\mathfrak{P}_\ast}$ is strongly connected since it is balanced.  From Lemma \ref{lemmamultiplicity} we further know that the multiplicity of the zero eigenvalue of ${L}(\mathscr{G}_{\mathfrak{P}_\ast})$ equals to the number of strongly connected components of ${L}(\mathscr{G}_{\mathfrak{P}_\ast})$. On the other hand $-{L}(\mathscr{G}_{\mathfrak{P}_\ast})$ is  the matrix representation of $\mathcal{K}_{\mathfrak{P}_\ast}$ under the basis $\mathscr{V}$. The desired conclusion holds.

\medskip

\noindent (iv). The conclusion follows from (iii) and the fact that ${\rm dim}\big({\rm Null}(\mathcal{K}_{\mathfrak{P}_\ast}) \big)$ is fully determined by $\mathfrak{C}_{\mathfrak{P}_\ast}$ from Lemma \ref{lemmanull}.

We have now completed the proof of Theorem \ref{propoperatorgraph}.

\end{document}